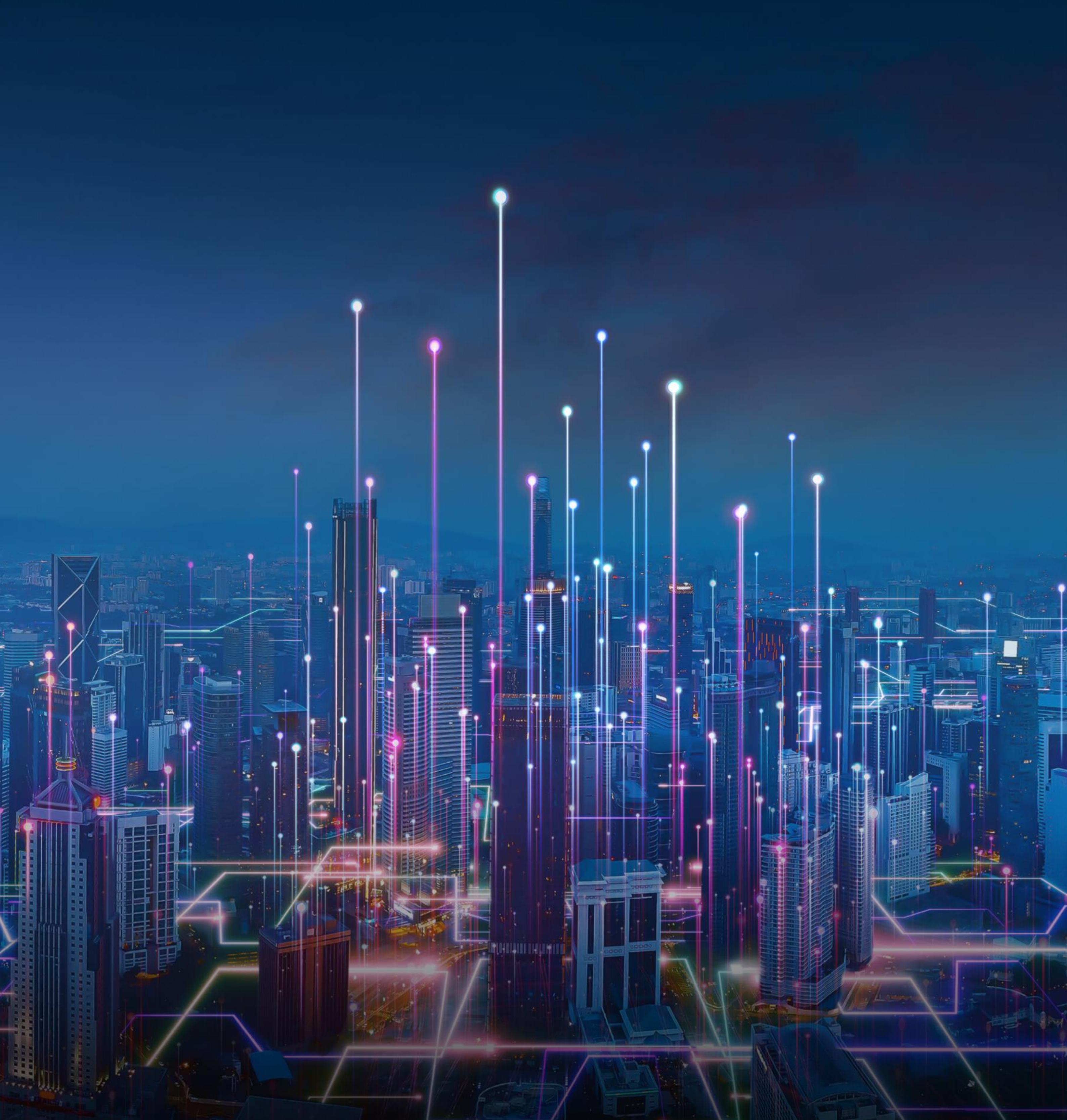

# 6G OSS 技术白皮书

联合发布:

AsiaInfo 亚信科技　AIR 清华大学智能产业研究院　中国电信 CHINA TELECOM　中国移动 China Mobile　中国联通 China unicom　CBN 中国广电　intel

## 联合作者

亚信科技

清华大学智能产业研究院

中国电信研究院

中国移动研究院

中国联通研究院

中广电移动网络有限公司

英特尔（中国）有限公司

## 引用本白皮书

《6G OSS 技术白皮书》，欧阳晔，张亚勤，叶晓舟，刘云新，王希栋，孙杰，刘洋，王首峰，边森，李赟等，2023 年 7 月。

# 目录

# 图目录

# 一. 前言

随着移动通信技术从 1G 到 5G 的持续演进，移动通信已渗透到生产、生活的方方面面。在 5G 时代，移动通信系统已经从传统通信业务提供者转变为促进千行百业数智化转型的赋能者。第 6 代移动通信系统（6G）作为新一代智能化综合性数字信息基础设施，将实现通信感知计算一体化、空天地海立体覆盖等能力的跃升，具备泛在互联、普惠智能、多维感知、全域覆盖、绿色低碳、内生安全等典型特征，实现从服务于人、人与物通信到支撑智能体高效联接的跃迁，全面引领经济社会数字化、智能化、绿色化转型。

业界普遍预期 6G 将在 2030 年实现商用[1]。各主要国家已陆续布局和启动了 6G 相关研究工作，全球 6G 技术竞争已经拉开序幕。截至 2022 年底，各大国际标准化组织、行业组织和研究机构相继发布了超过 30 本 6G 预研白皮书。国际电信联盟（ITU）、第三代合作伙伴计划（3GPP）等主要标准化组织面向 6G 的标准化工作有望在 2023 年启动。目前，业界对于 6G 的研究主要聚焦于愿景需求、典型业务场景、潜在关键技术和网络架构，但是对于网络必要组成的 6G 运营支撑系统（OSS）尚缺乏前瞻性、系统性的研究。

OSS 系统是移动通信网络的核心支撑。现有 5G OSS 系统实现了网络配置管理、业务编排、性能管理、故障管理、安全管理等重要功能，并针对上述功能引入了数据平台和人工智能以实现 5G 网络大数据处理和网络智能化[2]。6G 网络将实现空天地一体组网、通感算融合业务，需要 6G OSS 系统对包括地面蜂窝网、卫星网络、高空平台等多种物理网络和通感算多种资源的一体化运维管理，支持 6G 基础网络的技术演进。当前 5G OSS 系统的智能化能力水平，难以满足 6G 更加复杂、庞大的网络系统所需的“规、建、优、维、营”全生命周期管理要求，6G OSS 必须通过高水平的自动化、智能化和数字孪生化能力，实现端到端的网络全自智运维，支撑 6G 典型业务场景运营。此外 6G OSS 系统还需要在环境、社会、治理（ESG）领域发挥更大的社会责任，在面临紧急情况和灾难时充分发挥 6G

空天地一体化组网和通感算业务融合的优势，支撑 6G 成为绿色、高效、安全的全球数字基础设施。

本部白皮书基于亚信科技及各产学研伙伴在 5G OSS 系统上的研究和建设基础，通过对当前 6G 研究现状和 OSS 标准化进展的分析，提出了 6G 时代 OSS 的总体愿景，研究探索了 6G OSS 系统的潜在关键技术和功能架构，给出了从 5G OSS 向 6G OSS 系统的演进方案。希望为下一代移动通信的发展建设提供有益思路。

# 二. 全球 6G 研究现状

## 2.1 6G 愿景与需求

随着全球数字化转型的不断深入，移动通信网络深刻影响着人类的生产、生活。到 2030 年，社会服务均衡化、社会治理科学化、社会发展绿色化将成为未来社会的发展趋势，经济高质量发展、环境可持续发展等要求将驱动移动通信网络由通信基础设施向数字基础设施转变，推动 5G 万物互联到 6G 万物智联的发展。

中国 IMT-2030（6G）推进组提出"万物智联、数字孪生"[1]的 6G 总体愿景：6G 将与先进计算、大数据、人工智能、区块链等信息技术交叉融合，成为服务生活、赋能生产、绿色发展的基本要素；将充分利用低中高全频谱资源，实现空天地一体化的全球无缝覆盖；将提供完全沉浸式交互场景，支持精确的空间互动，满足人类在多重感官、甚至情感和意识层面的联通交互；通信感知和普惠智能不仅提升传统通信能力，也将助力实现真实环境中物理实体的数字化和智能化；将构建人机物智慧互联、智能体高效互通的新型网络，具备智慧内生、多维感知、数字孪生、安全内生等新功能；将实现物理世界人与人、人与物、物与物的高效智能互联，打造泛在精细、实时可信、有机整合的数字世界。

北美 Next G 联盟（Next G Alliance）6G 路线图报告提出 6G 愿景的 6 大目标[3]：强调在所有条件下的可信、安全和弹性；增强数字世界体验提升生活质量和创造更高经济价值；低成本高效能的解决方案；基于虚拟化技术的分布式云和通信系统增强动态性、提升性能和弹性；AI 内生的网络为应用提升鲁棒性、性能和效率；与能源效率和环境相关的可持续性，以实现到 2040 年 IMT 碳中和的目标。

由包括中国移动、美国蜂窝电信公司（US Cellular）和沃达丰公司（Vodafone）等全球主要运营商在内的下一代移动通信网络（NGMN）联盟发布的《6G 驱动力和愿景白皮书》[4]指出：引入新的人机界面，将用户体验扩展到多个物理和虚拟平台以满足各种使用情况；使用地面和非地面网络，提供跨陆地、海洋和天空的覆盖；确保在能源消耗和碳排放的严格限制下，以成本和能源效率提供具有极为多

样化要求的异构服务，以实现可持续性和碳中和的目标；确定适当的基于人工智能的框架，以支持价值创造和交付、资源分配优化、可持续部署和运营等。各大通信设备厂商也就 6G 网络愿景和需求发布了各自的观点。爱立信在其白皮书中提出 6G 发展的驱动力来源于对可信任网络、可持续发展、基于人工智能的便捷生活，以及探索新型未知应用的需求，其中，最典型的应用场景是数字、物理世界的信息交互[5]。华为在其《6G-无线通信新征程白皮书》[6]中则认为 6G 将跨越人联和物联，迈向万物智联，推动各垂直行业的全面数字化转型；三星的 6G 愿景是为人类和机器提供更高阶的连接体验，包括身临其境的 XR 服务，以及高保真和数字孪生服务[7]；而诺基亚认为 6G 将扩展和改变现有网络功能，融合人类、物理世界和数字世界，以释放我们与生俱来的人类潜力。[8]

全球学术界也积极参与 6G 愿景与需求的研究探讨。2019 年芬兰奥卢大学在其发布的《6G 泛在无线智能的关键驱动因素及其研究挑战》[9]首次将泛在无线智能作为 6G 的关键愿景。2020 年，英国萨里大学提出将支持物理世界和虚拟世界融合、实现无处不在的覆盖作为 6G 的新战略愿景[10]；同年，中国的东南大学联合上海科技大学、英国南安普敦大学等国内外科研院校联合发布了《6G 研究白皮书》[11]，提出“全覆盖、全频谱、全应用、强安全” 的 6G 无线通信网络的发展愿景：空天地海一体化网络用于提供深度全球覆盖；sub-6 GHz 频段、毫米波、太赫兹、光频段在内的全频谱资源充分挖掘以提供更高的数据传输速率；人工智能将与 6G 无线通信网络高效融合以实现更好地网络管理与自动化，并提高下一代网络的性能；包括物理层与网络层安全在内的强安全或内生安全。

综上所述，全球对 6G 愿景已形成基本共识。6G 将通过全频谱、全覆盖、安全可靠、绿色节能和普遍智慧，超越连接，实现网络空间与人类社会、物理世界、数字世界的深度融合。

6G 典型部署场景将分别具有高流量、高密度、高移动、高精度、高智能、广覆盖等特征。典型部署场景下的 6G 关键性能需求，将主要包含体验速率、峰值速率、流量密度、空口时延、同步和抖动、连接数密度、移动性、可靠性、覆盖、感知/定位精度、AI 服务精度等，满足 Gbps 体验速率、千万级连接、亚毫秒级时延、7 个 9 的高可靠、厘米级感知精度、超 90%智能精度等关键性能需求[12]。

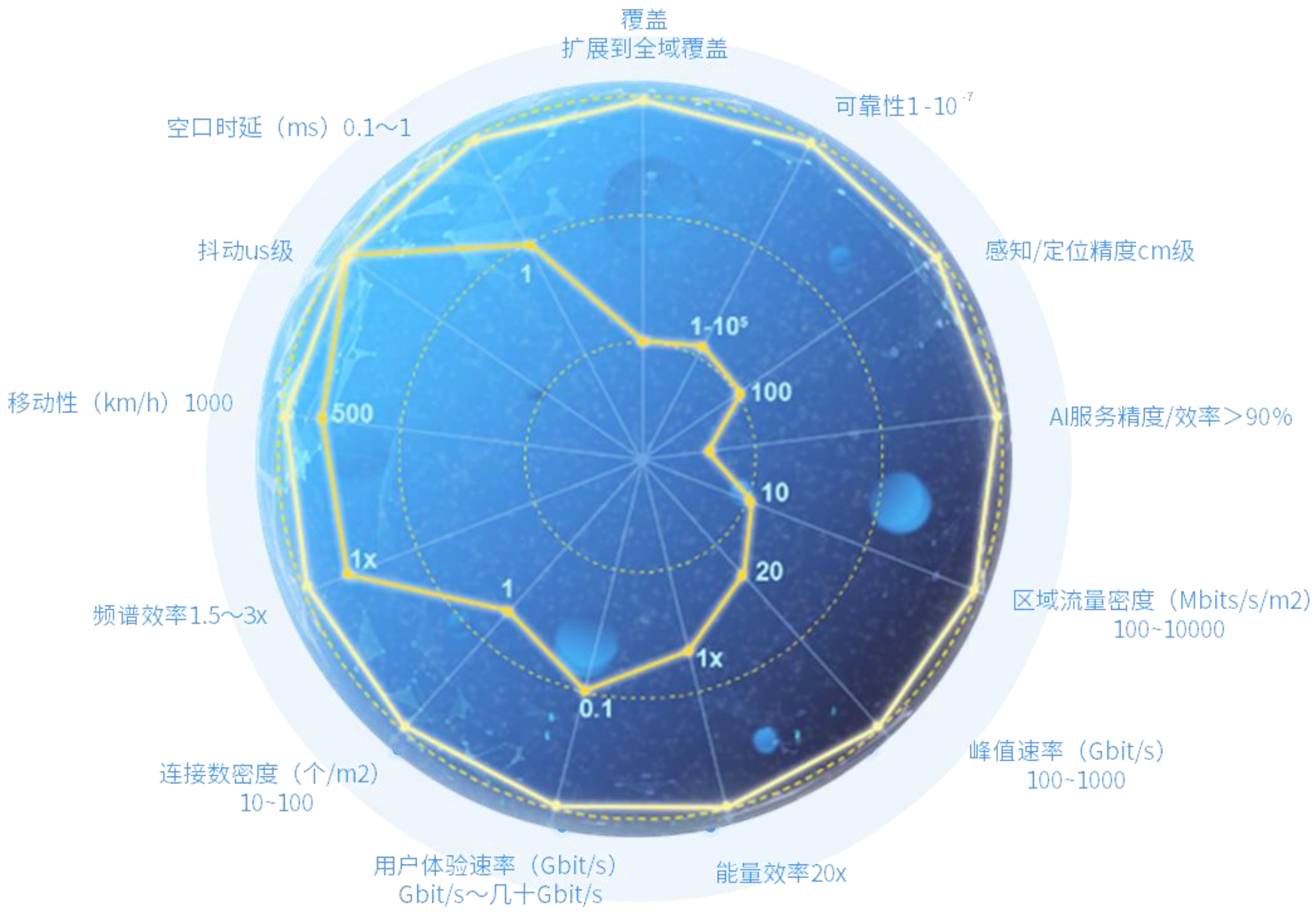

**图 2-1 6G 网络关键性能需求**

## 2.2 6G 典型业务场景

全球通信标准化及行业组织面向 6G 的典型场景的讨论已经逐步清晰，6G 将在 5G 增强移动宽带 eMBB，超可靠和低延迟通信 uRLLC 和大规模机器类型通信 mMTC 三大典型场景持续增强的基础上，进一步拓展面向新需求和新技术的新场景，其中普惠智能服务、感知通信融合等特性将被纳入 6G 的新场景中[13,14]。

超级无线宽带是增强型移动宽带（enhanced Mobile Broadband, eMBB）的演进和扩展，涵盖了更加多元的人机交互，不仅将极大提升以人为中心的沉浸式通信体验，也将在全球任意地点实现无缝覆盖。典型应用包括沉浸式 XR 和全息通信、远程多感官呈现及互联、工业机器人触觉反馈和控制。此外，语音服务的独立支持是沉浸式通信不可或缺的一部分。该场景将具有极高的数据速率，以及更低的延迟和更大的系统容量。它涵盖了从密集城市热点到农村的所有类型的部署。这些场景均对峰值速率、用户体验速率、系统容量、频谱效率提出更高的要求。由于环境数据的采样密度非常高，将使用网络上的可靠计算来负载计算复杂的处理和渲染，或实时远程访问渲染的图像。此外，6G 网络还需要提供低时延和高稳定性以保障用户体验。

极其可靠通信将在超可靠低时延通信（Ultra-Reliable and Low Latency Communication, URLLC）的基础上进一步增强能力，包含对传输可靠性和可用性具有非常严格要求的通信，如 IMT-2020 之后的极端 URLLC（时间敏感、信任等）。该场景还用于连接性以外的服务，例如可靠计算、精确定位或连接对象描述、数据分发、AI 原生 RAN 设计和其他网络平台功能，对数据速率、延迟、抖动灵敏度、功率限制、设备连接密度等其他特性的要求因所考虑的用例而异。对于某些应用，能够提供可预测的性能差异至关重要。典型应用包括机器人协作、无人机群和各种人机实时交互操作、自动驾驶、远程医疗手术、以及智慧能源、智能家居领域的应用等。此类应用普遍要求更低时延和更高可靠性，其中机器协同交互类的应用对抖动、时间同步、稳定性等确定性指标也提出了极高需求，因此需要网络同时具备中高速数据传输和超高精度定位的能力。

超大规模连接将在 5G 大规模机器类型通信（massive Machine Type Communication，mMTC）的基础上，拓展全新的应用领域和能力边界，重点是需要物联网和移动宽带连接能力，大量传感器不仅连接部署数量大，而且地理分布广泛，这将对覆盖范围提出很高的要求。此外，延长电池寿命、扩展覆盖范围和低成本也是需要考虑的关键因素。超大规模连接的对象将包括部署在智慧城市、智慧交通、智慧农业、智能制造等场景的各类设备和大量传感器，基于数字孪生技术，通过建模、推演、决策等环节与物理世界交互，可能需要支持高精度定位、高可靠和低延迟等能力。与 5G 中仅支持大规模设备的低速率传输相比，6G 超大规模连接设备的传输速率将从低到高不等，且业务需求差异化明显，需要网络提供多样灵活的性能支持。

普惠智能服务是 6G 新增典型场景，智能服务是未来 IMT 网络提供的新的超越通信服务，用于支持 AI 驱动的应用程序以及设备内 AI 功能。它的特点是将 AI 原生功能整合到未来的 IMT 网络和应用程序中，依托网络对需要进行高效分布式智能学习或推理的智能化服务提供集成化的通信和 AI 算力，由网络内生的大量智能体共同执行复杂的 AI 训练和推理任务，提高网络整体的性能和效率。普惠智能服务需要网络提供可靠的计算、分布式学习和推理能力、多功能性和可用性的保

障。此外，实现原生可信的网络安全和数据隐私保护也是该场景的重要目标与关键基础条件。

通信感知融合是 6G 新增典型场景，感知和通信的集成将提供高精度定位、环境重构、成像等多元化能力，极大促进超高分辨率和精度的应用需求，如超高精度定位、高分辨率实时无线地图构建、基于设备甚至无设备的被动目标定位、环境重建和监控、手势和动作识别等。这一场景增加了新的性能维度要求，例如对距离、速度、角度的感知分辨率、感知精度、检测概率等，其指标需求因应用而异。

## 2.3 6G 潜在关键技术及网络架构概览

为实现未来 6G 网络“人、物理世界、数字世界”智慧连接的美好愿景并满足极致的性能需求，当前各研究机构针对 6G 网络提出了 23 项潜在关键技术[1,3]。按照技术类型的不同可分为四类：

- **6G 无线技术**：太赫兹通信技术、可见光通信技术、动态频谱分享技术、超大规模 MIMO 技术、延迟多普勒域波形技术、先进调制编码技术、全双工技术、非正交多址接入技术、语义通信技术、智能超表面技术、全息无线电技术、轨道角动量技术。
- **6G 网络技术**：空天地一体组网技术、确定性网络技术、分布式自治网络技术、可编程网络技术、服务化 RAN 技术。
- **6G 融合技术**：通感一体化技术、网络内生 AI 技术、数字孪生网络技术、算力网络技术。
- **6G 安全技术**：内生安全技术、区块链无线接入网技术。

### 2.3.1 6G 潜在关键技术

#### 6G 无线技术

6G 无线技术包括太赫兹通信技术、可见光通信技术、延迟多普勒域波形技术、超大规模 MIMO 技术、智能超表面技术、全息无线电技术等。

太赫兹通信技术可作为现有空口传输方式的有益补充，将主要应用在全息通信、超大容量数据回传、短距超高速传输等潜在应用场景，同时借助太赫兹通信信号进行高精度定位和高分辨率感知也是重要应用方向[1]。

可见光通信指利用从 400THz 到 800THz 的超宽频谱的高速通信方式，具有无需授权、高保密、绿色和无电磁辐射的特点，可见光通信技术比较适合于室内的应用场景，可作为室内网络覆盖的有效补充[1]。

延迟多普勒域波形技术将信号的数字域处理和分析由时频域迁移到延迟多普勒域，利用延迟多普勒信道的稀疏性进行信号处理和分析，通过延迟多普勒域到时频域的转换获得发送信号的时频域分集增益对抗多普勒引起的载波间干扰，有望提升 6G 高移动性场景下的传输速率[7]。

超大规模 MIMO 技术是 MIMO 技术的进一步演进，天线和芯片集成度的不断提升推动天线阵列规模持续增大，超大规模 MIMO 技术可在更多样的频率范围内实现更高的频谱效率、更广更灵活的网络覆盖、更精细的定位精度和更高的能量效率。而且分布式超大规模 MIMO 有助于构造超大规模的天线阵列，使网络架构趋近于无蜂窝形式的无定形网络，促进实现均匀一致的用户体验，获得更高的频谱效率，降低系统的传输能耗[15]。

智能超表面技术（RIS）采用可编程新型亚波长二维超材料，以软件控制的方式对无线传播环境主动控制，在三维空间中实现信号传播方向调控、信号增强或干扰抑制，可应用于高频覆盖增强、克服局部空洞、提升小区边缘用户速率、绿色通信、辅助电磁环境感知和高精度定位等场景[7]。

智能全息无线电（IHR）是利用电磁波的全息干涉原理实现电磁空间的动态重构和实时精密调控，将实现从射频全息到光学全息的映射，具有超高分辨率的空间复用能力，主要应用场景包括超高容量和超低时延无线接入、海量物联网设备的高精度定位和精准无线供电以及数据传输等[14]。

除了以上五种技术外，语义通信技术、动态频谱分享技术、先进调制编码技术、全双工技术、非正交多址接入技术、轨道角动量技术等也都是 6G 潜在无线关键技术。此外，无线网络云化也是 6G 重要演进方向之一，可以满足 6G 无线

网络深度融合通信、感知、计算、人工智能等多样化能力的需求，是构建开放、灵活、高性能的 6G 无线网络，实现 6G 无线网络按需服务能力的重要技术基础。

### 6G 网络技术

6G 潜在网络关键技术主要包括空天地一体组网技术、服务化无线网技术和分布式自治网络技术等。

空天地一体组网技术将地面网络、不同轨道高度上的卫星（高中低轨卫星）以及不同空域飞行器等融合而成为星地一体的移动信息网络，通过地面网络实现城市热点常态化覆盖，利用天基、空基网络实现偏远地区、海上和空中按需覆盖，具有组网灵活、韧性抗毁等突出优势[15]。

服务化无线网技术将传统集成单体基站解耦为控制面和用户面服务，通过服务化接口实现功能服务之间的交互与能力开放，以按需组合的方式提供更灵活或更精简的网络服务能力，助力提升网络对全行业的适应能力[16]。

分布式自治网络技术包括接入网和核心网在内的 6G 网络体系架构，对于接入网，应设计旨在减少处理延迟的至简架构和按需能力的柔性架构，研究需求驱动的智能化控制机制及无线资源管理，引入软件化、服务化的设计理念；对于核心网，需要研究分布式、去中心化、自治化的网络机制来实现灵活、普适的组网。分布式自治的网络架构涉及去中心化和以用户为中心的控制和管理、需求驱动的轻量化接入网架构、智能化控制机制及无线资源管理设计等多方面关键技术[14]。

此外，确定性网络技术、可编程网络技术也是 6G 潜在关键网络技术。

### 6G 融合技术

目前，潜在的 6G 融合技术有四种，即通信感知一体化技术、网络内生 AI 技术、数字孪生网络技术和算力网络技术。

通信感知一体化技术的设计理念是要让无线通信和无线感知两个独立的功能在同一系统中实现且互惠互利。一方面，通信系统可以利用相同的频谱甚至复用硬件或信号处理模块完成不同类型的感知服务。另一方面，感知结果可用于辅助通信接入或管理，提高服务质量和通信效率。通信感知一体化技术通过收集和分

析经过散射、反射的通信信号获得环境物体的形态、材质、远近和移动性等基本特性，利用经典算法或 AI 算法，实现定位、成像等不同功能[14,17]。

网络内生 AI 技术将 AI 模型内生于移动通信系统并通过无线架构、无线数据、无线算法和无线应用等呈现出新的智能网络技术体系。6G 网络内生 AI 可分为内生智能的新型空口和内生智能的新型网络架构。内生智能的新型空口将打破现有无线空口模块化的设计框架，实现无线环境、资源、干扰、业务和用户等多维特性的深度挖掘和利用，实现网络的自主运行和自我演进；内生智能的新型网络架构利用网络节点的通信、计算和感知能力，通过分布式学习、群智式协同以及云边端一体化算法部署，使得 6G 网络原生支持各类 AI 应用，构建新的生态和以用户为中心的业务体验[18,19,20]。

数字孪生网络（DTN, Digital Twin Network）是一个具有物理网络实体及虚拟孪生体，且二者可进行实时交互映射的网络系统[21]。在此系统中，各种网络管理和应用可利用数字孪生技术构建的网络虚拟孪生体，基于数据和模型对物理网络进行高效的分析、诊断、仿真和控制。同时，数字孪生网络服务作为一种新的网络服务为业界提供端到端或部分网络功能的孪生服务，使能移动网络创新加速，以降低电信行业研发成本和缩短研发周期。数字孪生网络系统通过物理网络和数字网络实时交互数据，相互影响，可以帮助实现更加安全、智能、高效、可视化的智慧 6G 网络[22]。

算力网络技术将云边端多样的算力通过网络化的方式连接与协同，实现计算与网络的深度融合及协同感知，达到算力服务的按需调度和高效共享。算力网络的管控系统将由网络进一步向端侧延伸，通过网络层对应用层业务感知，建立端边云融合一体的新型网络架构，实现算力资源的无差别交付、自动化匹配，以及网络的智能化调度，并解决算力网络中多方协作关系和运营模式等问题[14]。

### 6G 安全技术

6G 内生安全技术的架构应奠定在一个更具包容性的信任模型基础之上，具备韧性且覆盖 6G 网络全生命周期，内生承载更健壮、更智慧、可扩展的安全机制，涉及多个安全技术方向。融合计算机网络、移动通信网络、卫星通信网络的 6G 安全体系架构及关键技术，支持安全内生、安全动态赋能；终端、边缘计算、

云计算和 6G 网络间的安全协同关键技术，支持异构融合网络的集中式、去中心化和第三方信任模式并存的多模信任架构；贴合 6G 无线通信特色的密码应用技术和密钥管理体系，如量子安全密码技术、逼近香农一次一密和密钥安全分发技术等；大规模数据流转的监测与隐私计算的理论与关键技术，高通量、高并发的数据加解密与签名验证，高吞吐量、易扩展、易管理，且具备安全隐私保障的区块链基础能力；拓扑高动态和信息广域共享的访问控制模型与机制，以及隔离与交换关键技术[17]。

区块链无线接入网（B-RAN）是一种由区块链技术支持的去中心化、可信任的无线接入范式。区块链无线接入网络在支持频谱共享、协作传输、多跳数据传输、设备对设备通信等的同时，可以在服务提供商和客户之间建立可信的物理链接。由于其分布式的特性，区块链无线接入网络能从本质上支持处于前沿的联邦学习，利用网络效应吸引更多的参与者，通过通信、计算、缓存和控制单元的集成和协调，向整个网络提供智能服务[23]。

## 2.3.2 6G 潜在网络架构

随着 6G 关键技术研究与探索的不断深入，全球 6G 推进组织均提出了 6G 网络架构的演进建议。

中国 IMT-2030 提出分布式自治的 6G 网络架构[[24]，如下图所示，具有分布式、定制化特点的6G网络架构不仅可以抵御DDoS攻击和降低单点故障的风险，也可以为每一个用户提供定制化的策略。去中心化的用户和数据管理方式，也让终端用户获得了个人数字资产的所有权和控制权，提供 DaaS 数据服务，结合智慧内生的网络 AI，提供 AIaaS 智能服务。同时，IMT-2030 提出 6G 智能内生网络体系框架，从下到上依次为异构资源层、功能和编排管理层以及能力开放层，旨在构建一张人机物智慧互联、智能体高效互通的智能网络，最终实现“万物智联，数字孪生”6G 总体愿景。

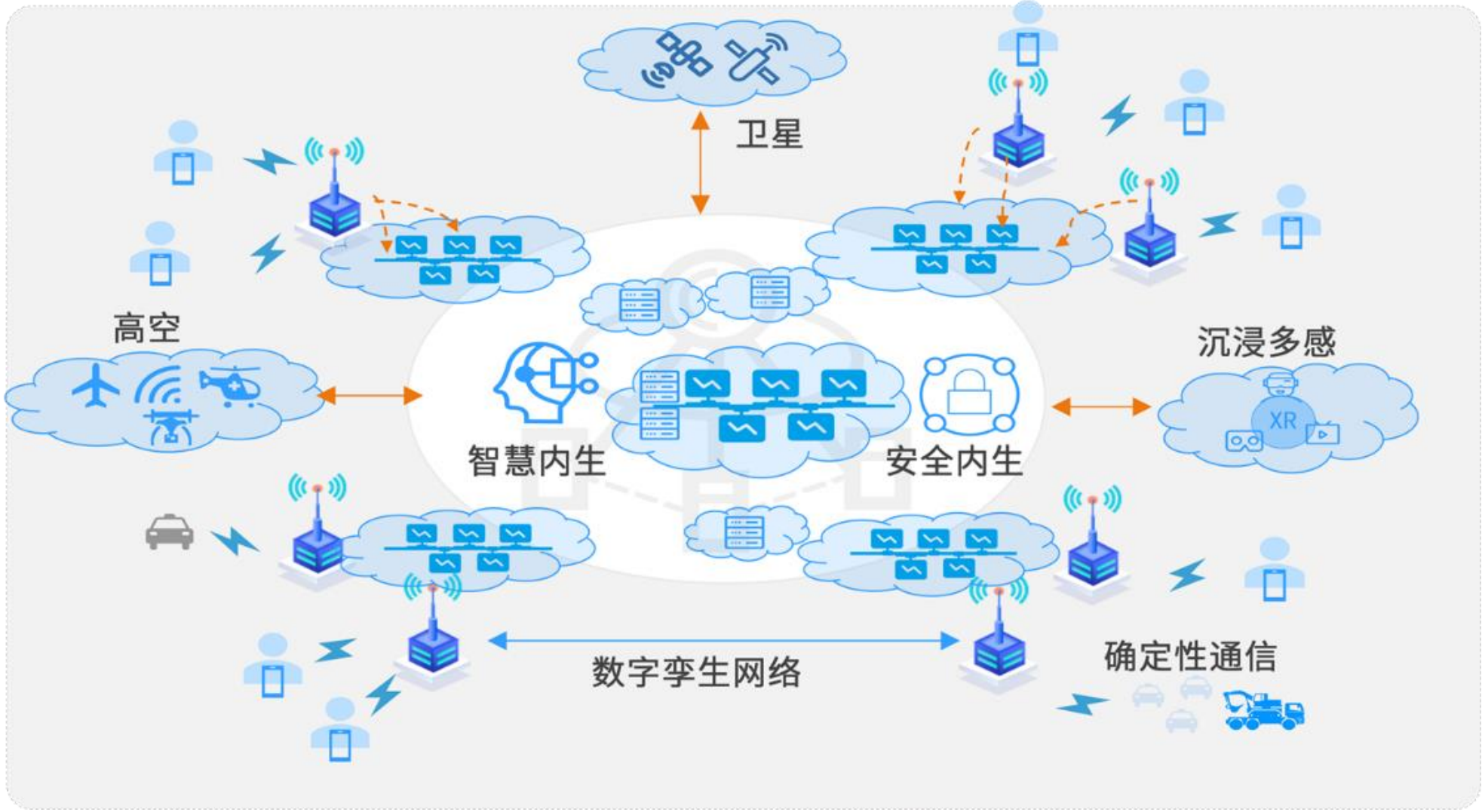

**图 2-2 分布式自治的 6G 网络架构愿景**

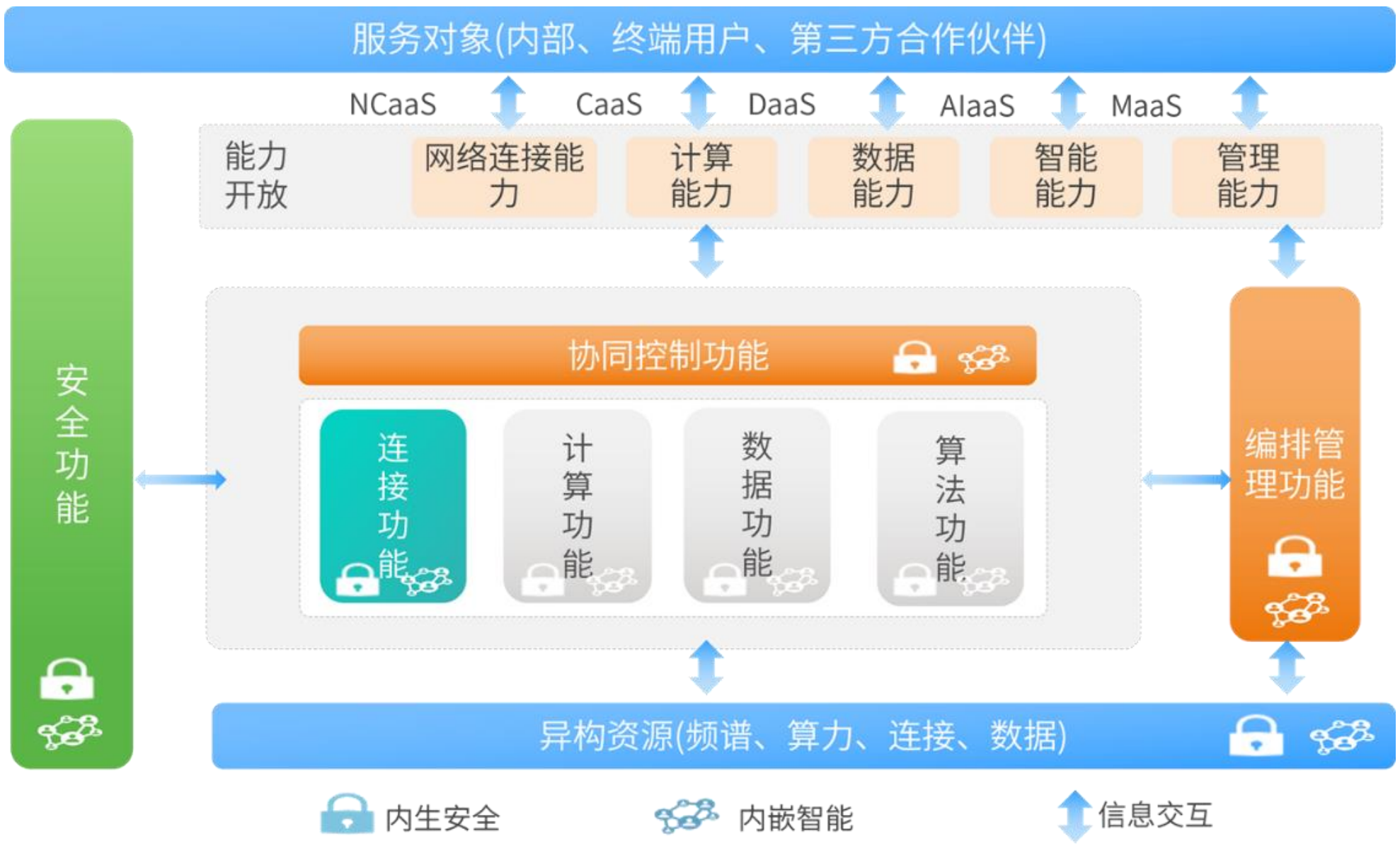

**图 2-3 6G 智慧内生网络体系框架**

欧盟 5G PPP Architecture Working Group 提出了包括基础设施、网络服务与应用的端到端 6G 网络架构，如下图所示，主要遵循能力开放、基于 AI 的自动化、灵活拓扑、可伸缩、弹性和可用性、服务化开放接口、网络功能解耦与简化原则。同时，欧盟启动 Hexa-X 6G 无线网络计划，并提出智能 6G 网络架构，强调 6G 网络的人工智能/机器学习与可编程等技术的应用[25,26]。

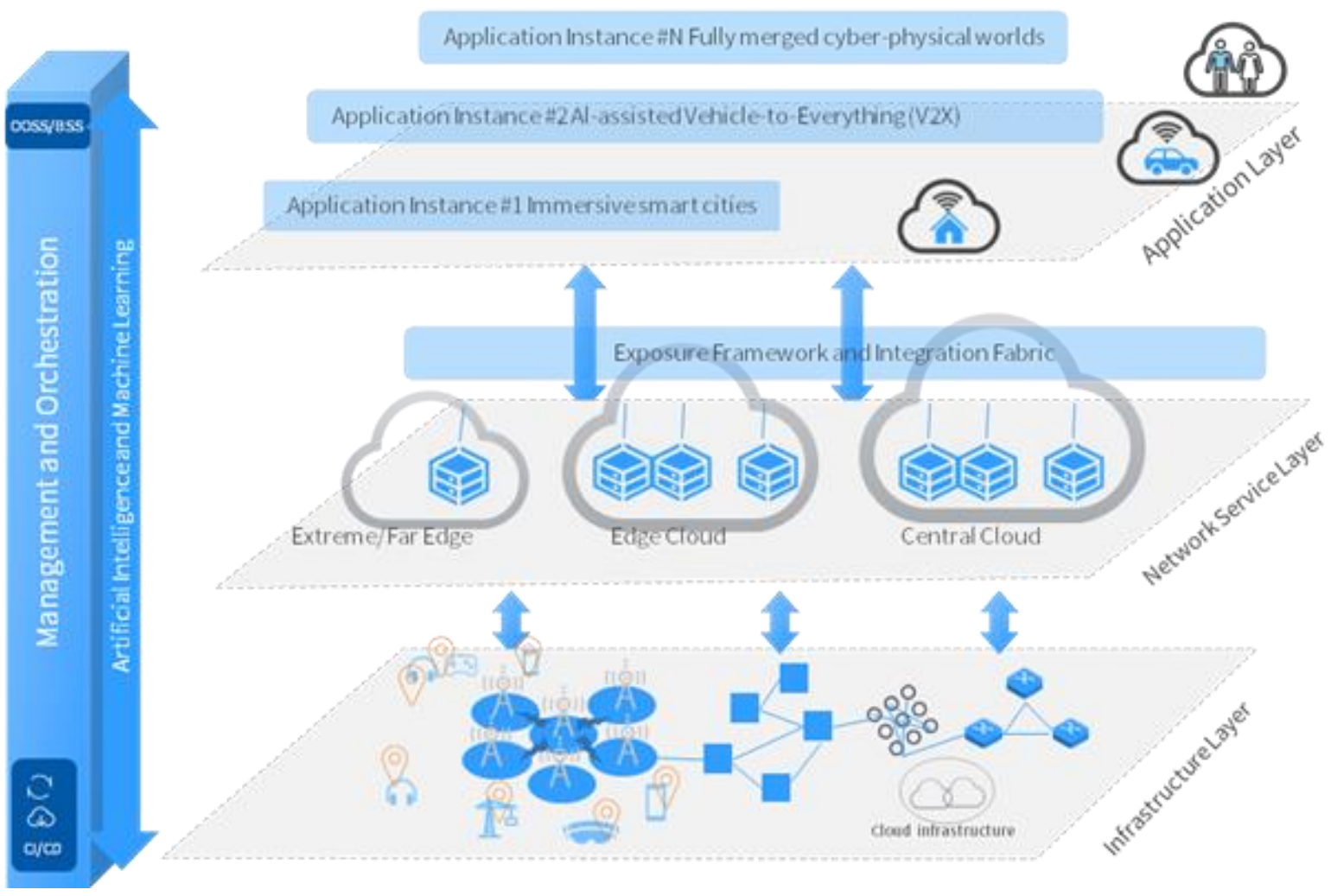

**图 2-4 欧盟 5G PPP 6G 网络架构**

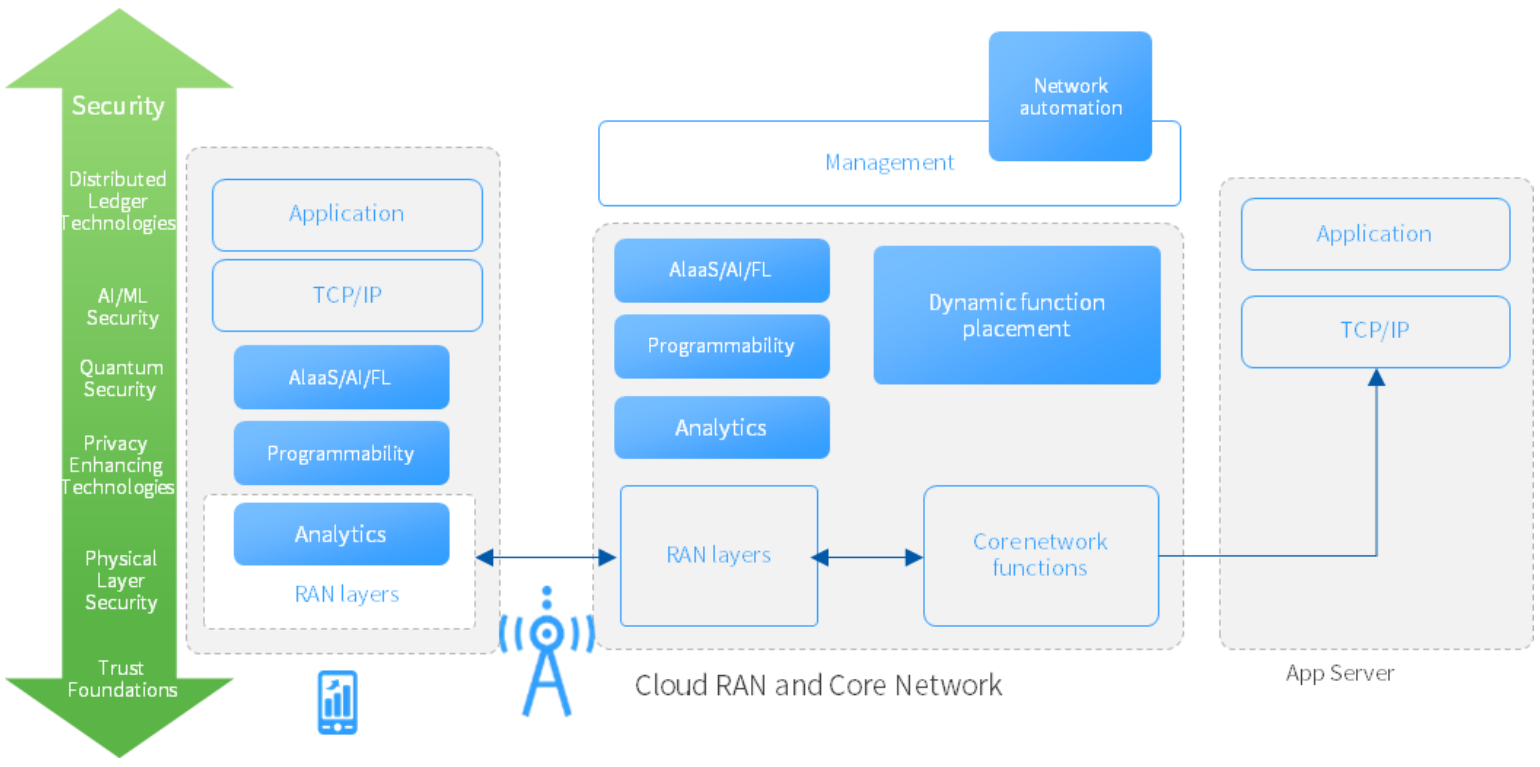

**图 2-5 欧盟 Hexa-X 智能 6G 网络架构**

美国电信行业解决方案联盟（ATIS）发起成立北美 Next G 联盟，推动北美在 6G 及未来移动技术方面的领导地位，并提出了非地面网络（Non-Terrestrial Networks，MTN）、网状和侧链 RAN 拓扑结构、基于服务与分布式 NAS(Non Access Stratum)的网络架构、网络解耦、分布式云平台、人工智能/机器学习在网络和设备中的应用等 6G 网络架构关键技术，加速推进北美市场的 6G 网络技术研发、部署和商用[27]。

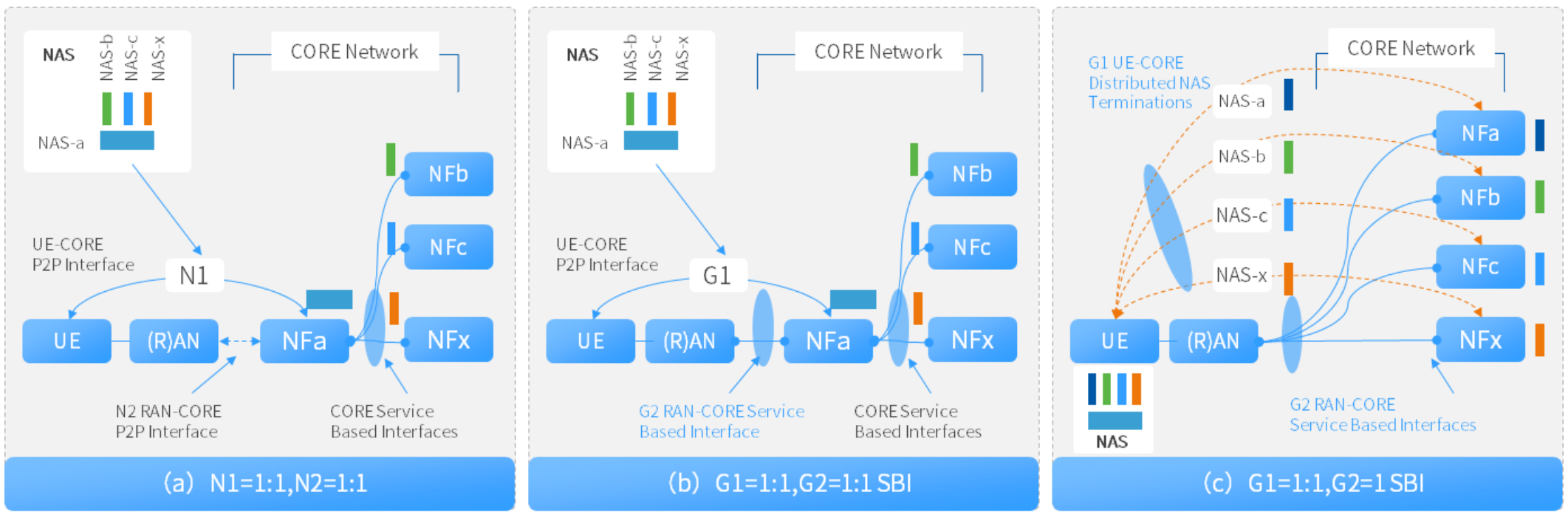

**图 2-6 Next G 基于服务与分布式 NAS 的网络架构**

除了各大 6G 推进组织，业界各主要运营商和设备厂商也各自提出了 6G 网络架构的研究思路。中国移动针对 6G 网络设计了“三体四层五面”的总体架构，包括网络本体、管理编排体、数字孪生体三大实体，网络本体的逻辑层又分为资源与算力、路由与连接、服务化功能、 开放使能“四层”，同时增强传统控制面、用户面功能并引入新的数据面、智能面、安全面，共同组成“五面”[16]。SK Telecom 与 NTT docomo 联合发布的《6G 共同需求》白皮书提出 Open RAN 将成为 6G 的默认形态以及网络云原生、网络自动化对 6G 的更高要求[28]。华为认为 6G 网络架构应该注重网络原始 AI、面向任务连接、多模信任架构和以用户为中心定制服务的设计范式[6]。爱立信对于 6G 网络架构的研究则侧重基于无蜂窝 MIMO 技术开发由“分布式无线电、计算和存储架构”组成的新型无线接入网络[5]。亚信科技研发了业界首套算力网络全栈软件基础设施产品[29,30]，并针对 6G 通感算业务一体化运营的特点进一步提出了算力内生网络架构[31]，可利用智能化通算调度编排决策机制，实现 6G 通算业务质量保证同时提升网络系统资源利用率，降低 6G 网络能耗。

## 2.4 6G 潜在关键技术和网络架构对 6G OSS 的影响

6G 潜在关键技术和网络架构不仅带来 6G 网络性能的提升，同时也对 6G OSS 的技术发展和架构演进提出了新的需求和挑战，可归纳为以下五个方面：

- **面向 6G 网络新架构和新技术的网络运营能力提升需求**

  空天地一体组网技术要求 6G OSS 系统需要支持空基、天基、地基网络的网络协作融合和空天地一体组网的网络编排。超大规模 MIMO、智能超

表面等技术将使无线网络通信业务相关参数数量大幅增长，对于 RIS 系统的动态调控以及 RIS 与 MIMO 结合的波束赋形为 6G OSS 中针对网络基础覆盖的规划和优化工作带来了新的可选维度。而服务化接入网和分布式自治网络技术也要求 6G OSS 需要支持 6G 核心网和接入网的全栈服务化架构和分布式组网架构的网络运维。因此，6G OSS 网络运营的广度、维度、深度均有更高要求，运营复杂度构成巨大挑战。

- **面向 6G 网络泛在智能的 OSS 内生能力需求**

  6G 网络将智慧内生、安全内生和数字孪生作为潜在关键技术，这也要求 6G OSS 系统需要具备网络智慧内生的端到端管理能力和 OSS 自身的内生 AI 能力、数字化孪生网络能力以支持网络规建优维的自智网络演进，同时 6G OSS 系统自身需要具备内生安全能力以确保网络数据安全和用户隐私并支持网络内生安全功能的运维支撑。如何实现由网络外挂式的人工智能、安全向内生转变，由网络仿真向数字孪生转变，是 6G OSS 必须解决的关键技术问题。

- **面向 6G 网络通感算一体化业务的编排管理需求**

  6G 网络通感算一体化要求网络同时具备物理-数字空间感知、泛在智能通信与计算能力，网络内的各网元设备通过通感算软硬件资源的协同与共享，实现多维感知、协作通信、智能计算功能的深度融合，这就需要 6G OSS 具备针对通感一体化业务、通算融合业务产生的通感算资源的编排与调度能力。一方面如何应对编排对象复杂度指数级增长的挑战，另一方面如何打通现有条块分割，真正形成一体化调度，将是 6G OSS 编排管理必须解决的关键问题。

- **面向 6G 网络对外赋能和社会责任的需求**

  IMT 2030 和 5G PPP Architecture Working Group 提出了 6G 网络能力灵活开放等特征，需要通过 6G OSS 对网络能力进行统一服务化的开放管理，实现 6G 网络能力按需灵活开放，从而为行业数智化转型赋能，同时 6G OSS 也需要助力 6G 网络在可持续发展和公共安全等方面更好的承担社会责任。

- **面向 6G 网络与新型 IT 技术融合的需求**

  6G 网络架构和业务类型的丰富使网络数据规模更加庞大，网络运维操作更加复杂，6G OSS 需要结合数据治理技术提升数据管理效率与数据应用价值，引入 RPA 技术全面实现 6G 网络规划、建设、维护、优化与运营的超级自动化，同时 6G OSS 还需支持 6G 网络向全栈服务化的演进。因此，6G OSS 必须面对更广泛技术革新的挑战。

# 三. OSS 相关网络管理标准化进展

2022 年 6 月 ITU-R WP5D 发布了 ITU 首份面向 2030 年及未来 IMT 无线技术发展趋势的研究报告《未来技术趋势研究报告》，内容涉及人工智能与无线通信的融合、感知通信融合、无线空口技术增强(如大规模天线、调制编码与多址接入、高精度定位等)、新维度无线通信(如智能超表面等)、太赫兹通信、无线网络架构等重点技术方向。此外，工作组正在编制《未来技术愿景建议书》，该建议书包含面向 2030 及未来的 IMT 系统整体目标，如应用场景、主要系统能力等。3GPP 也将在 2023 年 Rel-19 阶段开始 6G 愿景、技术、需求方面的工作，Rel-19 阶段不仅将定义 5G 系统的新增能力，还将为 6G 系统需要具备的能力提供指导。3GPP 预计将在 2025 年下半年开始对 6G 技术进行标准化（完成 6G 标准的时间点在 2028 年上半年），预计 2028 年下半年将会有 6G 设备产品面市。

OSS 的核心作用是保障网络性能并提升网络管理效率，网络运维及管理是各大标准化组织的重点标准化领域。多年来为适应通信网络在不同发展阶段的运维及管理需求，3GPP、ITU、TMF、ETSI 以及 O-RAN 均在持续推动与 OSS 相关的标准化工作。结合不同标准组织的工作范围以及协调配套的工作原则，ITU 作为联合国下设通信领域的权威国际标准化机构，重点关注网络运维管理整体架构以及原则的规范；3GPP 重点关注网络侧网元管理功能设计以及接口规范；ETSI 关注在网络虚拟化基础上的网络管理技术规范；O-RAN 则重点关注以开放 RAN 为基础的新型网络管理模式以及实现路径；TMF 作为电信运营和管理领域的权威专业性国际组织，系统性全面关注业务支撑和网络运营管理方面的技术探讨以及规范制定。

## 3.1 3GPP 网络管理标准

3GPP 网络管理方面的标准研究主要由 3GPP SA5 负责，包括网络的运营、管理、维护和供给保障（Operations, Administration, Maintenance and

Provisioning）。目前 3GPP SA5 在 Rel-18 阶段的标准研究重点为三个领域：网络智能化与网络自治、网络管理架构和机制、新业务支持[32]。

为支持网络智能化, 3GPP 在网络功能层引入网络数据分析功能（NWDAF），在管理层面引入管理数据分析服务（MDAS）。网络数据分析功能（NWDAF），基于标准化的服务接口，向 5GC/OAM/AF 提供按需、快速、精准的智能分析服务，支持多场景灵活部署，满足不同层级的应用要求，使能 5G 网络功能实体，实现运营商网络低成本、高效率的智能闭环。根据 Rel-18 阶段发布版本，NWDAF 进行了功能强化和解耦，将逻辑分析能力独立为 AnLF（Analytics Logical Function 分析逻辑功能）专用于数据分析，并可调用机器学习模型和能力、通过树形级联调取其他 NWDAF 数据联合分析，将机器学习模型和能力集中至 MTLF（Model Training Logical Function 模型训练逻辑功能）进行统一管理。此外，强化了数据采集作为独立功能 DCCF（Data Collection Coordination Function 数据收集与协调功能），并新增支持非 3GPP 标准化的信令框架（Message Framework）与 3GPP 标准接口 MFAF（Messaging Framework Adaptor Function 消息框架适配功能），并建立数据分析的存储管理 ADRF（Analytics Data Repository Function 分析结果与数据存储功能）。管理数据分析服务（MDAS）是 MDA 公开的服务，对管理域数据进行分析，支持 RAN 或 CN 域内的数据分析或跨域的数据分析，并支持与 NWDAF 的接口和交互，MDAS 可由各种消费者使用，例如 MNF（即网络和服务管理的 MnS 服务提供者/消费者）、NFs（例如 NWDAF）、SON 功能、网络和服务优化工具/功能、SLS 保证功能、人工操作员和 AFs 等。

3GPP 预计将在 2025 年下半年（Rel-20）开始对 6G 技术进行标准化（完成 6G 标准的时间预计将在点在 2028 年上半年），预计 2028 年下半年将会有 6G 设备产品面市[33]。可以预见在未来五年内，3GPP SA5 将在 6G OSS 相关领域针对空天地融合编排、通感算一体调度等方向开展新的标准课题研究。

## 3.2 ITU 网络管理标准

ITU-T （International Telecommunication Union-Telecommunication Standardization Sector, 国际电信联盟－电信标准部）自 1985 年开始制定的 TMN（Telecommunications Management Network）电信网络管理国际标准，是目前接受范围最为广泛的电信网络管理建设运营的基础标准。TMN 定义了两个网络之间的互连点，并指定了相关的网络管理功能，并先后发布了 TMN 系列建议书 M.3010、M.3400 和 X.700 等。TMN 的管理层次分为五层，从低到高依次为：网元层（NEL），网元管理层（EML），网络管理层（NML），业务管理层（SML）和事务管理层（BML）。其中网元层属于被管理层，其他四层属于管理层。作为 TMN 的补充，ITU-T 同时划分出了网络管理系统提供的五个通用的管理职能 FCAPS（Fault，Configuration，Accounting，Performance and Security，即错误、配置、计帐、性能和安全）。

目前，ITU-T 网络管理方面的标准研究主要涉及 ITU-T SG2 以及 ITU-T SG13 两个研究组。ITU-T SG2 针对 AI 增强的电信运营管理（AITOM）已提出了一系列标准项目的制定，ITU-T SG13 则分别面向自智网络以及网络 2030 启动了焦点组的研究工作[34,35,36,37]。

面向 2030，ITU-T SG13 研究组在 2018-2020 年成立了网络 2030 焦点组（FG NET-2030）探索面向 2030 年及以后的网络需求、使能技术及 IMT-2020（5G）IMT 系统的预期演进，探索面向 2030 年及以后的网络有望支持的新前瞻性场景，例如全息式通信、危急情况下的极速响应和新兴网络的高精度通信需求。同时，对于新兴需求和场景相适应的网络运维及管理的项目也将在现有架构基础上不断演进。

## 3.3 TMF 网络管理标准

TMF（Telecom Management Forum）是专门为电信运营和管理提供策略建议和实施方案的世界性组织,是专注于通信行业运营支撑系统(OSS)和管理问题的全球性的非赢利性社团联盟。自 1988 年成立以来，其领先的信息资源、知识

和技术方案被业界广泛认同。TMF 提出的 NG OSS（下一代运维系统）功能模型，包括了 eTOM、TAM、SID 和 TNA 四个框架模型，被国际电信运营商和设备制造商以及电信运营支撑系统开发商广泛接受，成为事实上的国际标准。面向未来数智驱动、泛在连接和虚实相生的通信行业新需求，TM Forum 推出了 ODA(Open Digital Architecture)开放数字框架。TMF ODA 用一种新的方法取代传统的运营和业务支持系统(OSS/BSS)来为电信行业构建软件，为标准化的云原生软件组件打开市场，并使通信服务提供商和供应商能够投资 IT 以实现新的和差异化服务[38]。TMF 为实现即插即用架构提供了一条进化路径，建立在 TMF 业务流程框架（eTOM）、信息框架（SID）、开放 API、数据分析和人工智能、客户体验管理和数字生态系统管理，通过开放 API 连接的标准化、可重复使用的软件定义组件，实现云端原生的即插即用 IT 和网络。

自动化方面，TMF 提出的 AIOps 服务管理是对现有 IT 框架（DevOps、Agile、ITIL 等）的演进补充，其中增加了具体的原则和做法，提出了需要在复杂的运营环境中采用并实施人工智能和传统应用的混合管理。AIOps 服务管理框架解决了在现有 CSP 的 IT 和网络运营中部署和整合大量 AI 组件及其相关业务能力所需的技术和运营流程。

智能化方面，TMF 自智网络项目中引入了意图，用来表达用户需求、目标和约束，允许系统相应地调整操作方式，与不同域的用户进行交互。在自智网络等级的中低层级（例如 L0~L3）中，用户需求、目标和约束可以使用策略驱动的操作和现有接口上承载的需求来实现。具有较高等级的自智网络（例如 L4~L5）系统将能够通过意图驱动的交互来自动调整行为，减少人工干预。这种能力将通过引入无需人工干预的、全新的、定制化的服务产品来提升业务灵活性[38]。

## 3.4 ETSI 网络管理标准

ETSI 中涉及网络运营管理的项目组主要包括 ETSI 零接触网络与业务管理工作组（Zero-touch network and Service Management，ZSM）、ETSI 网络功能虚

拟化工作组(Network Function Virtualization，NFV)以及 ETSI 体验式网络智能工作组（Experiential Networked Intelligence，ENI)。

ETSI ZSM 面向新兴和未来敏捷、高效可定性管理和自动化的网络管理，关注并定义水平和垂直端到端可操作的架构框架、解决方案以及核心技术实现。水平端到端指的是跨领域、跨技术方面。垂直端到端是指从面向资源的层到面向客户的层的跨层方面。目标是所有操作过程和任务（例如，交付、部署、配置、保证和优化）自动执行，理想情况下 100%自动化[39]。

ETSI NFV 通过构建 NFV 基础设施平台，提供支持托管的虚拟化网络功能的全生命周期独立部署和运营，旨在推动开放、可互操作的生态系统。NFV 定义了 MANO（网络功能虚拟化管理和编排）用于管理和协调虚拟化网络功能和支撑软件组件的架构框架，支持在虚拟机上部署与连接。自 2023 年起，NFV 将继续进一步整合技术规范并提供基于云原生技术、网络资源管理、网络管理和编排、网络连接技术、硬件和其他基础设施资源管理、虚拟化和云技术，以及新的用例（例如工业垂直行业和 vRAN）和运营模式的进步带来的新功能和运营需求[40]。

ETSI ENI 提供了基于模型、策略驱动、上下文感知的 ENI 系统实现对网络赋智。ENI 支持通过代理系统直接或间接的指定实体关联，指定实体包括 NMS、EMS、控制器，以及当前或未来的管理编排系统。ENI 系统基于体验式架构，通过自学习原则积累经验以持续提高运营效率，并使系统能够随着时间的推移实现提议到实施决策的全流程闭环控制。除了网络自动化之外，ENI 系统还协助人类和机器的决策，以实现更易于维护和可靠的系统，提供上下文感知服务，使运营商能够根据上下文变化调整服务[41]。

## 3.5 O-RAN 网络管理标准

O-RAN 联盟的目标是搭建一个开放、虚拟化和智能的无线接入网(RAN)体系结构，构建支持不同厂商和设备产品实现互操作的具有活力和竞争力的生态系统[42]。在传统网络架构的基础上，O-RAN 在网络边缘引入无线网络智能控制平台、无线智能管理器以及开放的标准化接口。

O-RAN 采用三层体系架构，从下往上分别为：云平台（O-Cloud）、O-RAN 网络功能以及服务管理和编排（SMO）框架。其中，SMO 的功能相当于传统封闭式的 RAN 接入网设备的网络运营和管理子系统 OAM 或 NMS。

网络智能化是 O-RAN 的四个核心方向之一，O-RAN 系统的网管与传统的封闭式网管的不同，体现为 O-RAN 运营的高度智能化、网络功能的服务化和可定制化，O-RAN 提供的网管不是传统的固化、面向单一厂家的网管，而是可以被调用的服务（Service），SMO 是多种管理服务的整合。在 O-RAN 体系结构中，SMO 主要负责 RAN 域的管理，比如：O-RAN 网络功能的 FCAPS 接口、O-Cloud 管理编排和工作流管理等功能及接口，以及用于 1s 以上 RAN 优化非实时智能控制环路的 Non-Real Rime RIC 功能及接口。

# 四. 6G OSS 总体愿景

基于 6G 愿景与需求，结合 OSS 系统对 6G 网络的核心支撑作用，6G OSS 的总体愿景可归纳为五个方面：

**实现由网络单体/单域管理到空天地、通感算一体化管理**

现有 5G OSS 系统支撑移动通信网络自身及通信资源的管理，为 5G 网络全生命周期提供智能化运维保障能力。6G OSS 系统将从对移动通信网络和通信资源的管理，扩展至对包括地面蜂窝网、高轨卫星网络、中低轨卫星网络、高空平台、无人机在内的网络空间，以及包括通信、感知、计算等多维度资源的一体化管理，通过立体化的管理能力提供无处不在的通信、计算、感知支持，提升 6G 系统的资源和能源利用率。

**实现从网络智能化管理扩展到网络自动化、智能化、数字孪生化管理**

网络与人工智能作为两个通用目的技术，在 5G 网络运维管理中，通过网络智能化提升网络运营效率。未来面向空天地、通感算一体化的复杂 6G 系统，6G OSS 需要构建自动化、智能化和数字孪生化等三化的全面能力，满足 6G 规、建、优、维、营全生命周期运维管理要求，提供前所未有的 6G 服务数字化体验。

**从 5G 自智网络 L5 级向 6G OSS ready 演进**

当前 5G 网络自智水平正在向中高等级（L3~L4）发展，目标是实现最高等级（L5 级），即具备面向多业务、多领域、全生命周期的移动通信网络全场景闭环自治能力。6G OSS 将在高等级自智的基础上，基于通感算一体实现智慧内生，并向全面自动化、高等级智慧内生和网络数字孪生的深度融合演进，扩展空天地一体自智范畴，实现 6G 万物智联、数字孪生的愿景。

**将环境、社会、治理（ESG）纳入 6G OSS 能力体系**

随着 5G 网络的建设发展，网络能耗问题已经凸显。现有移动通信网络作为信息基础设施，在面向 ESG 方面可提供一定的信息保障，但更多是 ESG 对其的外在要求。6G 作为新型的信息基础设施，需要在设计阶段就将 ESG 纳入 6G OSS 能力体系，基于空天地一体化、通感算一体化等全域覆盖能力和绿色低碳的网络

建设，确保 6G 通信系统履行更多的 ESG 和公共安全责任，支撑实现碳达峰和碳中和目标。

**构建安全可信的 6G OSS 体系**

高安全性是 6G 网络的重要特性,需要构建网络空间内生安全发展的新范式。6G OSS 作为 6G 网络的核心管理支撑面，要充分利用自动化、智能化、数字孪生化能力和网络编排管理调度能力，通过构建内生安全的 OSS 体系，为 6G 网络和应用提供服务与保障。

# 五. 6G OSS 技术框架与关键技术

## 5.1 6G OSS 技术研究思路

6G OSS 系统是面向未来网络的网络运维管理系统，需要从网络需求、系统演进和技术方向等多方面综合分析形成其关键技术。如图 5-1 所示，6G OSS 系统的技术研究思路主要由四方面组成：

- 支持 6G 网络新技术、新架构、新业务等所需的网络运维管理关键技术；
- 基于现有 OSS 标准化网络运维管理系统升级和功能扩展所需的关键技术；
- 支持网络智能化和自智网络演进的 OSS 网络智能运维关键技术；
- 将新型 IT 技术应用于 OSS 系统所需的 ICT 融合关键技术。

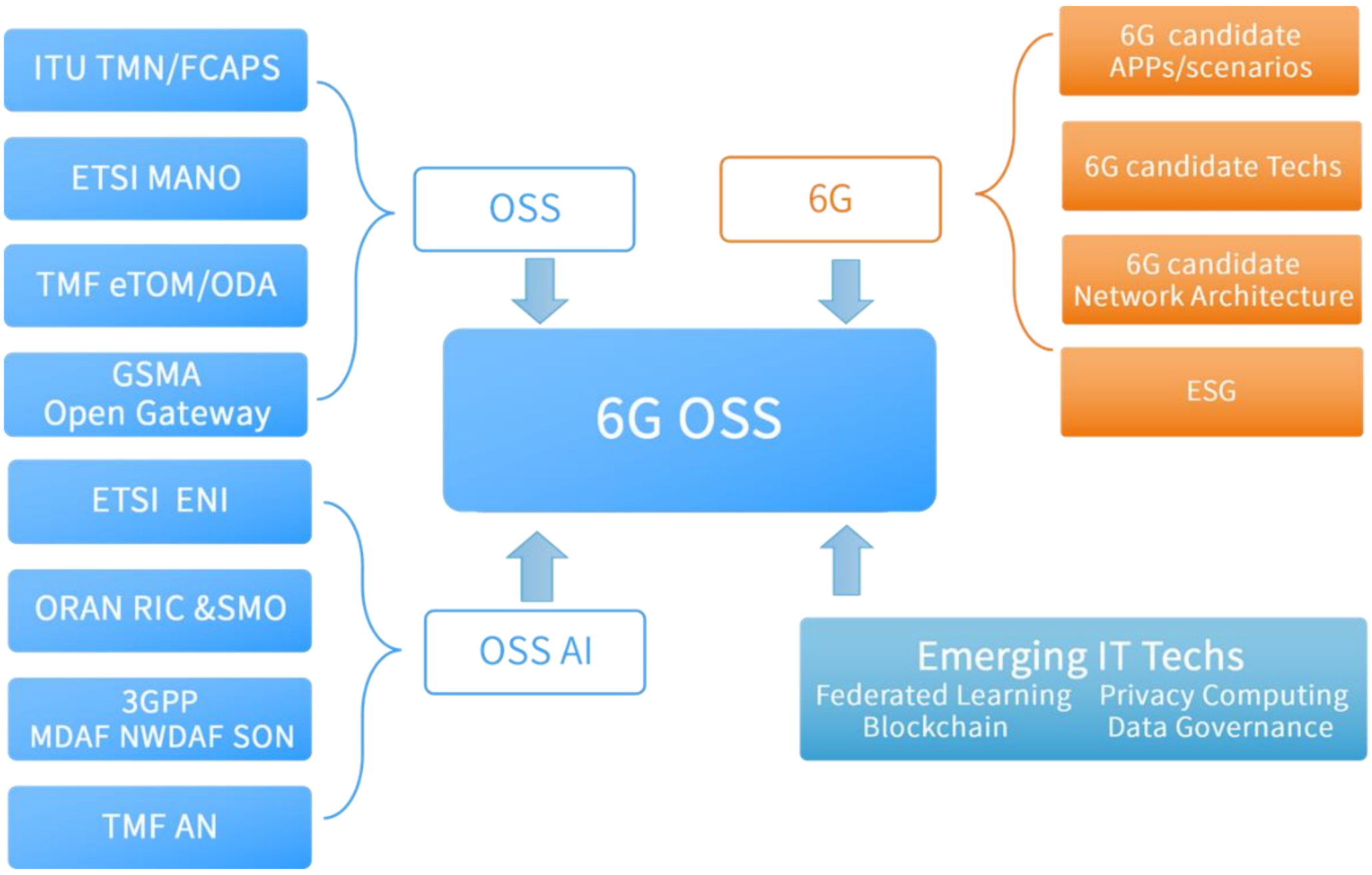

**图 5-1 6G OSS 技术研究思路**

# 5.2 6G OSS 的 12 项潜在关键技术

## 5.2.1 6G OSS 空天地一体化的网络编排

5G OSS 系统网络编排实现了 5G 业务的自动化开通。6G OSS 系统将支撑空天地一体组网，实现空基、天基、地基网络的深度融合，以及网络、计算、存储等资源的统一供给，在任何地点、任何时间、以任何方式提供信息服务。因此，6G OSS 系统网络编排针对空天地一体化要求，在实现基于虚拟网络功能的业务横向拉通之外，还需要实现空天地一体的网络域纵向拉通，以及通信、计算、感知等多种资源的弹性调度，从而满足空天地一体网络业务要求。如图 5-2 所示，由于 6G 空天地节点在通信性能、覆盖范围、链路质量等方面存在显著差异，一体化组网具有立体性、多样性、时变性、可扩展性等特征，6G OSS 网络编排需要进行空天地网络功能、通感算资源的立体式部署和弹性调度，实现空天地网络融合与优势互补，提高网络编排的灵活性、有效性和时效性，以适应不同应用场景需求和业务 SLA 要求。

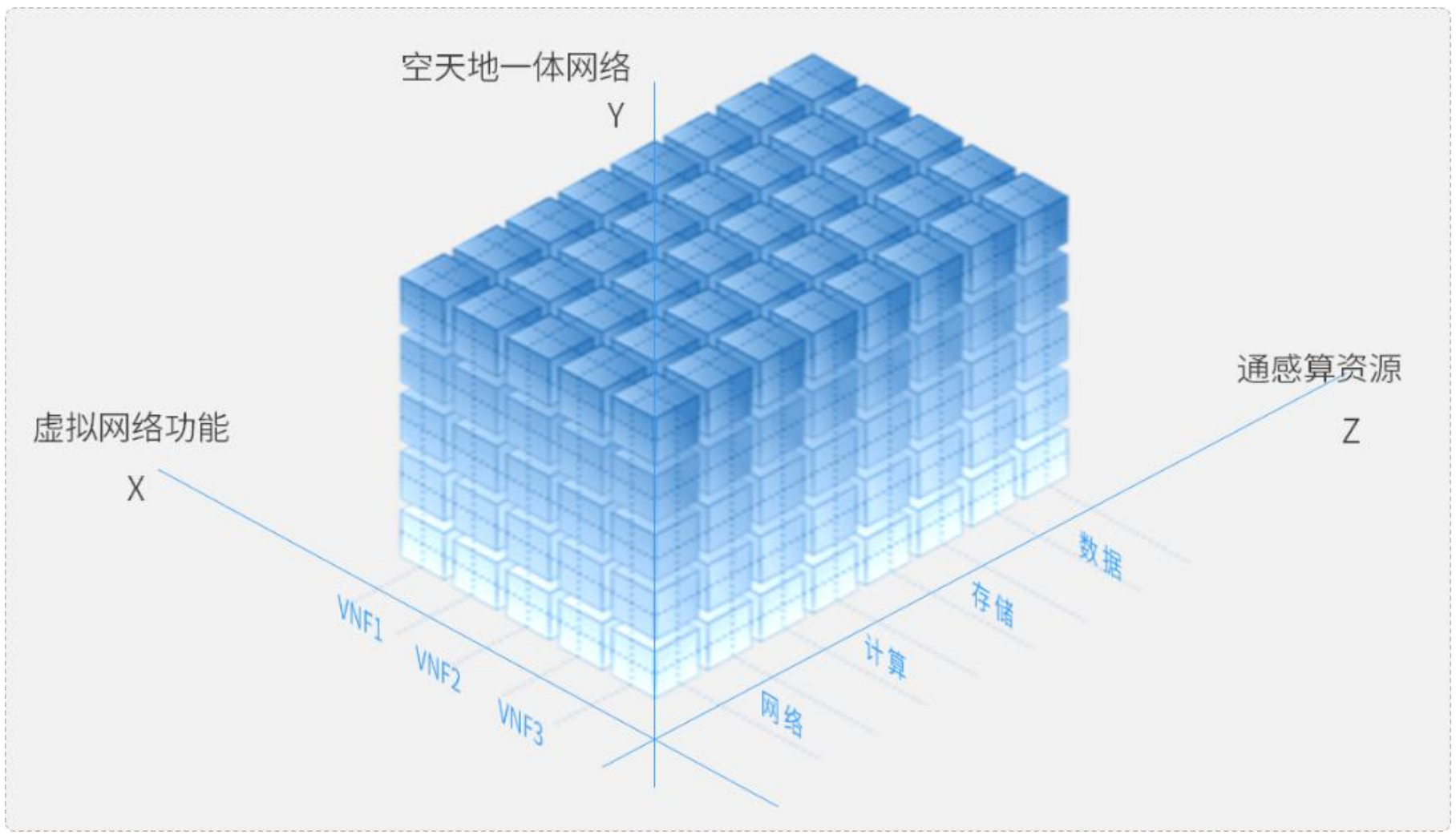

**图 5-2 6G OSS 立体化弹性网络编排**

6G OSS 网络编排工作流程：通过实时或者非实时方式采集空天地网络数据，包括各类网络资源、空间信息、配置参数、网络协议、接口、路由、信令、流程、性能、告警、日志、状态等数据信息；基于意图网络技术实现用户业务需求的准确理解；基于业务理解、采集的数据、策略规则、操作手册、专家经验等，通过

数据+知识驱动的 AI 能力，认知网络实时状态，预测网络状态走向、网络故障发生和定位、业务需求变化等，形成面向业务 SLA 的网络编排决策，实现对空天地网络功能、多维度资源的端到端编排与管控。

### 增强型意图网络技术

对于用户意图的准确理解是网络编排满足业务需求的前提条件。用户通过语音、文本等自然语言进行意图输入，意图网络技术将用户意图输入转化为网络意图表达模型并进行策略方案设计。在 6G 空天地复杂网络环境中，意图网络既要实现对用户意图的网络表达转译，也要实现网络策略模版的准确填充、意图和策略的验证，这就需要 6G OSS 具备增强型意图网络技术。该技术将以 GPT-4 为代表的多模态预训练大模型为引擎，网络管理人员仅需通过输入简单的运维意图并通过多轮交互的方式，就可较为高效的获知不同要求的网络运维策略，并根据网络策略下发后探测获得网络数据，检验意图达成情况。同时，脑机接口作为一种全新的控制和交流方式，在面向未来的科技创新发展中占有重要地位，在增强型意图网络技术中应用脑机接口也将逐渐成为可能。基于脑机接口，用户不再通过语音、文本进行间接的意图表达，而是直接通过大脑意识进行直接表达，并进而实现更加准确高效的网络表达转移。

### 基于网络遥测的实时数据采集技术

空天地网络时变性、动态性强，对数据采集实时性要求高。网络遥测技术从物理网元或者虚拟网元上远程实时高速采集数据，构建标准数据模型，支持一次订阅持续上报，采集数据的精度高，类型丰富，可以充分反映网络状况。网络遥测技术支持 OSS 系统管理更多的设备，数据采集过程对网络自身功能和性能影响小，可实现对网络实时、高速和更精细的监控，为网络编排提供大数据基础。

### 智能编排技术

空天地一体化网络编排的对象是多种功能类型和资源类型的网络节点，将网络设备的行为、能力进行功能抽象和软件化形成虚拟网络功能，将节点的网络、计算、存储、数据等多维度资源虚拟化，设计可编排的最小服务单元和多维度资源。智能编排将虚拟网络功能横向拉通，形成面向业务的网络功能集合；纵向拉通空天地一体化网络，实现通信、计算、感知等多维度资源的弹性调度。智能编

排提供统一的节点能力评价体系，节点能力由弹性的资源进行承载，当节点资源消耗超过安全边界时，节点性能可能呈现断崖式下降，需要保障弹性资源调度满足节点性能安全要求；基于用户意图分析，进行业务 SLA 需求与空天地网络功能和多维度资源的需求映射，如所需的端到端链路带宽；面向动态变化的网络全局拓扑，基于性能、资源、经济、绿色、安全等多方面因子进行综合分析决策，实现基于 AI 模型输出当前最优网络编排策略，指导网络功能编排与资源管控执行，选择保障用户业务 SLA 的网络节点和安全资源；进行相关策略编排后的执行效果评估，并基于编排策略带来的网络状态变化数据进行网络知识更新，实现 AI 模型的持续训练、持续部署和编排策略的持续优化。

### 5.2.2 6G OSS 对 6G 新无线技术的管理

#### 新形态网元的管理技术

6G 网络引入的新形态网元使 OSS 管理的对象属性更复杂。空天地一体化带来的天基、海基网元新增属性，不仅需要 OSS 对应建立属性参数的记录，更需要 OSS 根据新网元属性给予对应的管理和管控。对于天基网元，包括卫星系统构型、卫星载荷管理、星间通信链路管理、馈线链路管理等。由于 RIS 智能超表面带来的无线信道环境重构是需要与网络进行协同来实现的，6G OSS 需要具备管控 RIS 表面的能力。对于 RIS 表面的管理，需要增加其特有的反射特性参数控制能力、RIS 表面的位置管理等功能。对于复杂部署环境下通过 RIS 增加环境多径传输的场景，还需要联合基站与 RIS 表面进行信道的测量和反馈。为支持通信感知能力，6G OSS 将支持各类型的传感器的管理功能。传感器种类众多，对于不同感知能力对应的感知数据属性存在很大差异，传感器在网络中部署的位置和环境需要适配具体环境。对传感器生命周期状态的管理也将是 6G OSS 多元化网元管理中的新能力之一。面对更加智能化的物联网通信能力，6G OSS 还应具备相应的机器智能的管理和控制能力。例如对参与物联网智能决策可调用的 AI 模型管理、控制多种类物联网设备之间的信息和信息模型共享、决策生成管理等。应对智能物联网，6G OSS 将配合未来物联网组网技术演进，协同演进相应的控制和管理能力。

### 新无线中的通信感知数据管理技术

面向通信感知一体化技术，OSS 对应增加对感知数据的存储。OSS 管理的数据域维度扩张，感知数据的管理、感知信息的利用，成为新的 OSS 数据域管理内容。6G OSS 数据域的管理将向数据信息的管理演进。得益于内生智能对 OSS 的支持，6G OSS 对数据的处理将增加更高维度的数据信息融合能力。5G 网络运行数据、网络业务感知数据、无线环境感知数据、网络意图推导的配置策略等构成了 6G OSS 数据域管理集合。通过内生 AI 能力，OSS 将上述高维信息进行模型化并生成网络调整策略，引导网络自智调整和优化资源配置、性能表现。

### 智能化数据流向与路由管理技术

OSS 管理的数据流向控制更为复杂。伴随着通信感知技术、空天地一体化技术，以及去小区化技术引入 6G 系统，6G 网络内的数据流向更为复杂，数据分流和路由管理、业务 QoS、QoE 保障的数据分流选项更加复杂多样。同时，伴随着 AI 内生的成熟，OSS 内部也将演进出现面对数据分流及路由管理的 AI 能力。

面向通信感知，OSS 需要将感知数据处理为感知信息，并将感知信息与网络资源、业务质量要求结合，用于最大化满足系统性能同时保证业务质量要求的网络资源编排策略生成。面向空天地一体化，OSS 需处理好卫星转发与地面系统内路由之间的融合调度关系。对于单一业务的数据路径，考虑地面光纤传输经多个节点转发，或通过一跳卫星链路转发之间的时延差异，通信链路质量区别。对于网络局部的整体业务满足能力和服务质量保证，还受到每一跳转发的容量限制，制定单一业务转发时需要合理预估路径各节点容量的制约。MIMO 技术发展引导向大规模分布式天线联合传输，这一技术演进将改变 5G 网络的蜂窝小区结构设计。在 6G OSS 中，面向去小区化的网络结构，需要具备对天线收发对（一根发送天线到一个接收天线形成的一对收发天线关系）或由多个物理天线形成的虚拟天线对的数据发送关系的管理能力。在多用户环境下，分布式多天线与各用户的通信链路连接关系必须经过合理的设计，满足不同用户业务传输需求的同时，控制对其他用户数据链路的干扰。

### Cell-free 模式下的新型网络优化

由于网络去小区化，在相应的网络部署区域，其网络质量评估 KPI 的计算方式，网络管理对象，以及运维方式都将显著区别于传统蜂窝小区组网形态下的网络质量管理和运维。首先，网络质量评估 KPI 的获取由扇区级别下沉至天线级别，未来分布式 MIMO 的天线规模巨大，网络直接统计的天线级别数据量级呈几何倍数增长。其次，网络管理的对象下沉为收发天线对，由于数据业务时变性和用户移动产生的空间动态特征，带来收发天线对的关系是时间和空间二维变化的，网络管理难度骤增。第三，去小区化的区域内，网络的维护不仅需要考虑天线物理部署位置、天线与 DU 的物理连接关系，还需要考虑多用户接入环境下天线收发旁瓣与周边天线方位关系，并高度依赖于基于数字孪生模拟的多天线干扰分析以快速迭代合理的天线调整方案及波束赋形方案。

**新组网模式下的全频谱协同管理技术**

6G 网络采用的多项新技术，如 MIMO 增强技术、非正交多址、全双工技术、空地系统频率空间复用等，在提升频谱利用效率的同时均伴随着网络局部区域内更加高效的网络协同管理需求。基站侧的调度配合 CU/DU 内生 AI 能力可以解决基站内各小区间的资源协调问题。在移动通信网络中，覆盖能力的分析还包括对小区簇及更大范围的连片区域的覆盖效果评估，通常连片区域定义为在地理空间上连续分布的、无线信号覆盖场景具有明显共性特征的空间，例如高校园区、地铁站、办公楼、居民小区等。对于这些连续覆盖区域，通常基站级的资源管理无法有效解决网络资源合理分配的问题。通过分布式 OSS 的资源管理及其内生 AI 能力，可以在分布式 OSS 管控的较大范围内完成局部最优的干扰管理和资源协同，从而有效提升 6G 网络多种新技术协调共存，为局部网络提供局部最优的无线资源利用方案。在分布式 OSS 管控交界区域，可通过分布式 OSS 间协同实现相对简单的干扰协调，满足边界区域的业务质量要求。

### 5.2.3 6G OSS 能力开放

网络能力开放有助于提升运营商网络价值，并带来用户体验的提升，随着 SDN, NFV，云原生等技术的发展，网络能力开放已成为 5G 时代的重要关键技术之一。5G 网络在能力开放方面，3GPP 标准引入 NEF（Network Exposure

Function，网络开放功能），由 NEF 提供 5G 网络能力的汇聚及对外开放，包括对内面向其他 5GC（5G Core Network，5G 核心网络）网元以及对外面向 AF（Application Function，应用功能）开放，如事件监控能力、QoS 能力与参数配置等能力的开放。

6G 时代，随着 DOICT 技术的融合发展，6G 网络将持续走向开放，进一步丰富对外开放的信息和能力。为了更好地支撑 6G 新业务场景，6G OSS 需要实现全面能力管理能力开放，通过对通信、感知与计算等能力的采集、编排与调度，为自有业务和第三方应用提供服务。基于当前工业界与学术界的面向 6G 网络潜在关键技术与架构的研究，建议 6G OSS 能力开放主要包括以下关键技术。

### 基于 GSMA Open Gateway 的能力开放技术

6G 普惠智能服务、通感算一体等新型数字化业务场景对网络能力开放提出了标准化、全连接的需求，为此 GSAM 提出了 Open Gateway 技术框架，旨在为应用开发者提供对运营商网络的通用访问，实现电信业设计和提供服务方式的范式转变，帮助开发者和云供应商通过单点接入全球最大的连接平台，在运营商网络中更快地增强和部署服务。随着基于 Open Gateway 的 API 与服务数量的增加，这就需要 6G OSS 具备这些 API 的管理与调度能力，主要包括：

- API 全生命周期管理：实现 API 的上架、编辑与下架管理
- API 订购管理：实现客户基于网络 SLA、开放能力等需求的快速订购
- API 调度管理：实现 API 及所需资源的调度管理
- API 运维管理：实现 API 运行状态、资源状态等的监控运维管理

从而高效支撑构建 6G Open Gateway 能力开放体系，为用户提供所需 6G 网络业务能力的标准化、全连接开放。

### 基于 TMF Open API 的能力开放技术

面向通信感知融合、及其可靠通信等数字化新业务场景， 6G OSS 需要依照 CRISP-DM 知识发现过程模型，构建 6G OSS API 体系（如图 5-3 所示），提供数据处理服务、模型训练服务、模型发布服务、模型部署预测服务等基础服务能

力，同时参照 TMF Open APIs 框架对外提供权限管理服务、资源管控服务、状态监控服务和系统安全服务，以保障系统高可用。

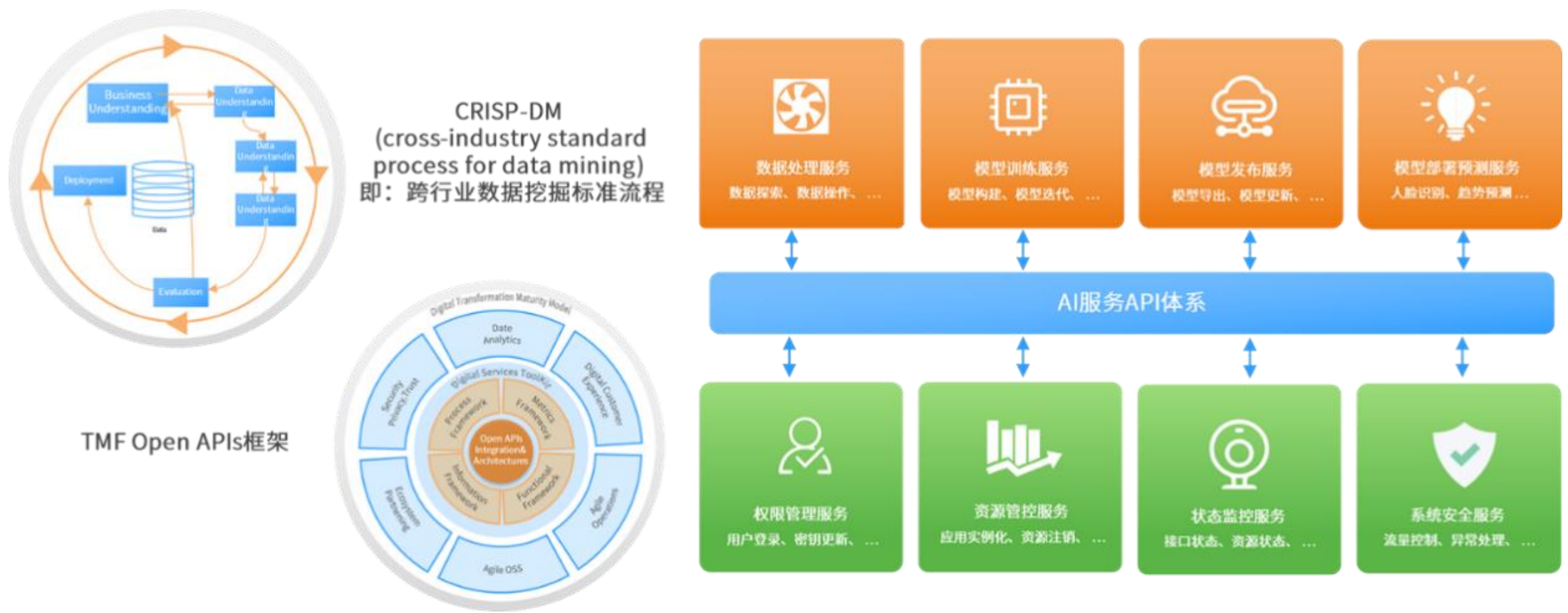

**图 5-3 6G OSS API 体系**

### 数字化能力开放技术

以用户体验为中心已经成为 6G 网络的目标愿景之一，同样 6G OSS 也需利用数字化技术(例如，数字孪生、语音识别等)，为运营商和消费者开放 6G OSS 能力和 6G 服务前所未有的数字化体验。面向 6G 专网运维运营体验提升，提供如基于计算机视觉的接口与 IT 故障排除，基于物联网传感器和生物识别的用户管理等技术能力。面向业务应用体验提升，提供如基于语音识别的服务，基于机器人的 QoE 提升与基于 AR/VR/MR/XR 的 QoE 提升等能力。

### 基于隐私计算、区块链的数据开放技术

新增数据面的 6G 网络架构将产生更多的价值数据，一方面，面向 6G 分布式网络架构，6G OSS 需实现网各分布式节点的数据自动采集，并增强标准化网络数据开放空能与接口，同时基于区块链技术保障数据的安全；另一方面，面向业务对数据的需求，需 6G OSS 需基于隐私计算实现 6G 数据管理，提升数据的价值流通与开放能力。

### 基于网络可编程的能力开放技术

5G 时代已经开始进行了一些网络可编程的探索和改进，如基于 SBA 架构的 5G 核心网控制面基于云原生的软件设计，使得 5G 核心网的控制面网络功能可以快速构建、发布及部署，结合云计算实现网络功能与底层硬件及操作系统解耦等。6G 时代，可编程技术将从控制面可编程向用户面可编程演进。服务化控制面

的网络功能支持容器化、云原生的方式部署，采用网络控制器，将配置下发到用户面。6G OSS 基于网络的可编程能力，将为用户提供更多的服务定制和更快的响应速度，主要包括：

- **灵活快速网络业务部署**：6G OSS 基于网络可编程能力，可实现网络控制面、用户面与数据面等逻辑、功能的灵活定义，减少冗余设计，统一网络服务能力，以更高的敏捷性、灵活性提供创新的通信服务，并支持业务的更快速部署。
- **智能场景开发**：基于不同的部署场景或用例，6G OSS 通过网络可编程能力实现网络能力的灵活定制化，并将人工智能引入网络服务设计和部署实施中，以更快速获取网络能力升级。

### 5.2.4 6G OSS 对 6G 全栈 SBA 服务化的支持

全栈服务化架构（SBA）是 6G 网络架构演进的重要方向之一，全栈 SBA 是服务化架构在接入网和核心网用户面领域的拓展；同时全栈 SBA 是服务化架构的进一步深化，从服务框架、服务接口、原子服务等方面增强，适应网络的分布式组织、服务的智能化调度、行业专网的灵活化部署。

为了支持 6G 全栈 SBA 服务化，6G OSS 需要从以下三方面进行技术演进：

**支持服务化架构在网络领域中的扩展**

6G 网络的服务化将在多个网络领域进一步拓展，但是对于不同网络领域的服务化扩展形式却不尽相同。在核心网领域，管理面的服务化设计形式与控制面的服务化不同，管理面的功能网元将内嵌一系列 MnS 或由一系列 MnS 集合而成的 MnF，同时不同管理功能网元间通过标准的 API 接口实现交互；用户面网元也将可能实现服务化，这将有助于 UPF 网元为更多的网元提供更加基础的 UPF 服务，但是 UPF 的服务化接口有可能是 gRPC 协议。在无线网领域同样需要实现服务化结构，一方面核心网与 RAN 之间需要定义服务化接口，另一方面用户面与物理层之间也需要定义服务化接口以实现用户面和物理层的解耦，促进多种底层接入技术在物理层的融合。

随着 6G 网络服务化的深入，6G OSS 将成为全栈服务化架构中主要的业务编排和服务控制载体，其对于无线网和核心网网元的运维管理方式将从传统的网元管理模式演进为微服务动态管理模式，通过 API 接口与各网络领域的微服务接口对接。

**支持服务化接口协议的演进**

在 5G 核心网的服务化架构中，3GPP 通过 HTTP2 协议与 TLS/TCP（Transport Layer Security/Transport Control protocol）协议结合的方式实现了 RESTful 架构特征的 SBA 接口。但是，由于 TLS/TCP 协议带来了接口安全性和传输速率的不足，在 6G 全栈 SBA 服务化架构的演进中也可能采用面向 HTTP/3 标准的 QUIC 协议，以改进 TCP 的拥塞控制机制，有效降低了连接建立的时间。因此，为支持全栈 SBA 服务化，6G OSS 需要支持基于 HTTP2 与 TLS/TCP 结合、QUIC 等多种协议的 SBA 服务化接口。

**支持 IP 传输协议的演进**

传统 IP 传输协议的设计原则是统计复用、尽力而为，无法满足 6G 网络全栈服务化的需求，而新型的 IP 协议将支持在用户面数据传输的过程中根据业务需求定制数据传输要求实现确定性业务。新型 IP 协议的演进主要在三个方向，一是更灵活的地址空间；二是在数据包格式中新增业务合约信息，该信息可携带业务类型、时延要求、丢包率要求等；三是数据包载荷的定制化。IP 传输协议的演进要求，6G OSS 需要具备根据新协议的包格式定义进行数据解析和交互的能力，以满足 6G 全栈 SBA 服务化需求。

### 5.2.5 6G OSS 数据治理

6G 网络数据相比 5G 呈现出更加海量、多态、时序、关联的特点，而 6G 智慧内生等都是基于数据驱动的自动决策， 因此需通过数据治理技术，保障 6G 网络数据质量，提升数据管理效率与数据应用价值成为 6G OSS 的关键技术之一。

5G 时代，5G OSS 主要通过网络数据平台/中台实现 5G 网络资源、告警、性能、质量等数据的集中化采集，统一处理与共享，并提供集中式多样化的数据服务支撑应用数据需求，提供 AI 注智与服务管控能力，支持智能化分析与通用能

力开放。其中，面向 5G 网络数据的数据治理，5G OSS 主要通过数据规划、标准定义、模型设计、数据开发、数据采集、数据创建、数据使用、数据归档、数据销毁数据治理“九部法”指导治理活动的开展，提供治理相关工具，对 5G 网络数据开展数据治理。

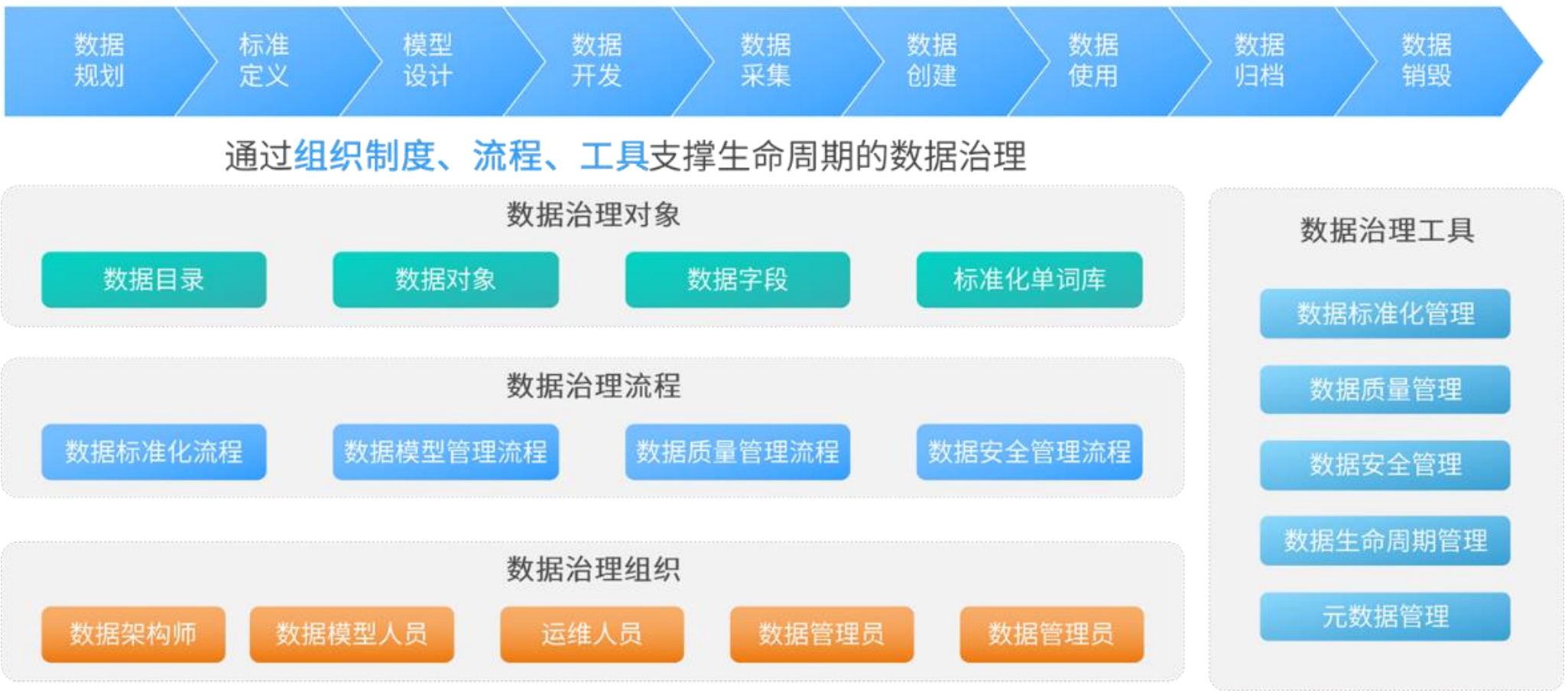

**图 5-4 6G OSS 数据治理流程**

6G 时代，基于 6G 网络数据海量、多态、时序、关联的特点，6G OSS 需要持续基于数据治理“九部法”，进行 6G 网络治理。如图 5-4 所示，治理流程主要包括：

### 制定数据治理标准

制定 6G 网络数据治理标准，从组织、流程、工具等方面，实现 6G 网络数据的标准化有效据有效治理。一方面，6G OSS 数据治理的范围，将从 5G 的网络资源、告警、性能与质量等数据数据，新增 6G 网络算力、存力与具备 AI 或分析能力网络功能（NF）产生的数据，如 NWDAF 、PCF 等；另一方面，随着 6G 数据治理范围的变化，需要对新增算力、存力与网络实时分析等数据制定相应的治理标准。包括数据分类与编码、数据字典、元数据标准、数据交换标准、数据质量标准等。

### 智能数据治理

基于机器学习和自然语言处理技术对 6G 数据进行分析，快速整理高频词根并将数据标准与元数据自动映射，建立数据标准管理体系；通过联邦学习来加强数据治理能力，提高数据交易数据流通的效率；通过机器学习自动识别数据质量，对数据质量进行效果评估和智能修复，并根据数据量和业务阶段的变化进行动态

更新；建立起业务部门与系统之间、多环节业务流程的信息采集、关联和交互，提高数据要素流通效率和精确度。

同时，6G OSS 将基于 DIKW 体系，通过知识图谱-机器学习等技术，首先基于 6G 移动通信原理与移动通信协议，将移动通信网络、算力与存力等数据进行全面梳理，完成面向人工智能的移动通信数据治理的第一步，即，使用专家知识库以及数据挖掘技术将专家知识和数据整合成人类所能理解的信息乃至知识规则。其次，利用知识图谱这项技术，构建具备逻辑推理关系的框架、模型，进行知识挖掘与知识推理。在网络数据知识图谱构建完成后，便可以对网络、算力与存力等进行全面监控。最后，机器学习作为闭环环节，主要是为了完成各类面向不同应用场景的特征数据集构建。

### 5.2.6 6G OSS 超级自动化

超级自动化是一种以业务为驱动、由多种技术构成的技术组合。面向通信行业 OSS 领域，主要基于机器人自动化(RPA)、流程挖掘（Process Mining）/业务流程管理(BPM)与低代码/无代码技术，实现网络规划、建设、维护、优化、运营，以及网络故障管理（Fault Management）、性能管理（Performance Management）、资源管理（Resource Management） 与配置管理（Configuration Management）的自动化与自治化。

#### 机器人流程自动化（RPA）

机器人流程自动化技术，主要用于信息化系统的自动化操作，在对现有系统非侵入的方式下提供数据采集、数据搬运、数据填写、流程执行、注智赋能、流程再造等能力，实现业务流程自动化、智能化，激发企业业务内生活力，助力企业数智化转型。

6G 时代，随着 RPA 技术的演进，6G OSS 将基于 RPA 技术全面实现 6G 网络规划、建设、维护、优化与运营的自动化。基于 RPA 技术，在网络规划领域，6G OSS 将实现无线网络建站选址、基站配置参数等的自动规划；在网络建设领域， 6G OSS 将实现新入网 6G 网元的自动信息采集录入等；在网络维护领域，6G OSS 将实现网络自动化巡检、各类运维数据的自动核查、网络故障与质量自

动分析等；在网络优化领域，6G OSS 将实现无线网络配置参数自动优化、通感融合参数自动优化等；在网络运营领域，6G OSS 将实现运营数据自动核查、业务方案自动上架等。

### 流程挖掘（Process Mining）/业务流程管理（BPM）

流程挖掘是数据科学（data science）和流程科学（process science）的粘合剂，其核心是从现代信息系统的事件日志中获得数据和提取知识，发现、监测和改进实际流程。BPM(Business Process Management)是一系列用于分析流程当前状态，设计“未来流程”以解决问题，并实施部署和监控流程，以改进流程的技术组合。通过流程挖掘技术提供的流程发现（Process Discovery）等能力，为 BPM 实现流程分析、设计、部署和监控提供强大支持。

6G 时代，随着流程挖掘（Process Mining）/业务流程管理(BPM)技术的演进，6G OSS 在 6G 网络网络故障管理、性能管理、资源管理与配置管理等领域将全面实现相关业务流程端到端自动化发现与设计管理。在故障管理领域，6G OSS 将基于故障预警与恢复流程进行挖掘与设计优化，实现网络故障从监控、预警、优化/恢复的流程端到端自动化；在网络性能管理领域，6G OSS 将基于网络性能优化流程行挖掘与设计优化，实现网络性能优化流程的端到端自动化；在网络资源管理领域，6G OSS 将实现 6G 网元从入网、在网到退网的全生命周期业务管理流程的端到端自动化；在网络配置管理领域，6G OSS 将实现网络配置参数从采集、优化到激活业务流程的端到端自动化。

### 低代码（LowCode）/无代码（NoCode）

低代码是一种用于快速设计和开发软件系统的技术，通过在可视化设计器中，以拖拽的方式快速构建应用程序，实现跳过基础架构以及繁杂的技术细节，直接进入与业务需求紧密相关的开发工作。无代码技术使非技术人员也可以使用它们来构建和部署自己的应用程序，而无需编写任何代码。

6G 时代，基于低代码（LowCode）/无代码（NoCode）技术，6G OSS 将支撑用户实现业务需求的快速落地与交付。在开放及管理方面，面向运营网络维护用户，6G OSS 可实现网络故障、性能、资源、配置等运维管理需求的可拖拽开发及配置；面向政企用户，基于低代码技术，6G OSS 可实现基于 6G 网络的新

业务快速开发。在扩展与集成方面，6G OSS 在快速可视化、可拖拽开发的基础上，通过少量代码或零代码，扩展现有组件功能以及通过集成外部能力来构建 6G OSS 能力。在用户体验提升方面，6G OSS 基于低代码技术向用户提供灵活、快速、便捷的 6G 网络与业务管理工具，降低用户学习成本，提升用户体验。在应用生态方面，6G OSS 基于超级自动化技术，实现 6G 网络与 6G OSS 应用的快速二次开发与能力开放，构建 6G 与 6G OSS 应用生态。

### 5.2.7 6G OSS 的 ESG 应用

6G 是实现社会服务均衡化、高端化，社会治理科学化、精准化，社会发展绿色化、节能化的重要基础设施，同时绿色与可持续性也是 6G 网络本身的目标愿景。6G OSS 支撑 6G ESG（Environmental 环境，Social 社会，Corporate Governance 治理）体系建设，赋能 6G 可持续和高质量发展。

**基于空天地一体保障的可持续性公共安全**

面对公共安全事件，如自然灾害、疫情、突发安全事件等，6G OSS 系统基于事前、事中、事后阶段的可持续公共安全网络保障与管理能力，实现网络通信快速自我重新配置与恢复，实现网络资源的动态调度，发挥 6G 网络更加广泛的公共安全社会责任。

基于对空天地一体组网、网络通算存多维度资源等动态网络数据实时采集分析，以及外部安全事件和态势联动，6G OSS 通过网络全生命周期自智技术，进行网络态势实时动态感知和预测，实现网络态势感知分析和预测预警；基于网络的故障分析和根因定位，实现故障的快速发现和快速定位，并通过自动化、智能化技术实现通过故障自动乃至无中断的快速恢复；对于安全事件物理损坏造成的局域性网络故障，可基于分布式、服务式专网技术快速恢复特定区域与范围的 6G 网络，或通过空天地立体组网优势，通过智能化网络编排调度其它网络资源，实现弹性组网和弹性资源调度，进行实时的通信能力应急恢复。

网络仿真与故障恢复演练是提高紧急公共安全情况下的网络应变能力,自我防护和恢复能力的重要方面。6G OSS 可基于数字孪生、内生 AI 等技术，针对各

类公共安全事件，实现网络故障发生、故障影响与故障恢复等端到端网络仿真与故障恢复演练的能力。

**基于空天地一体和通感一体的自然灾害防控**

世界自然灾害种类多，分布地域广，发生频率高，造成损失重。6G 网络空天地一体化组网，同时具有通感一体化能力。通过“空天地”一体化全域覆盖的网络，有效支撑灾害事故监测系统的信息感知、数据通信和服务承载，增大对灾害事故的监测覆盖面，提高精准度、时效性，增强决策指挥的科学性和有效性。利用 6G 本身通感一体能力，还可实现对已有可见光、红外遥感等成像能力的补充增强，提供灾害感知、辨识能力，识别灾害事故影响范围、发展动态、损毁情况等关键要素。

**绿色计算技术**

建立以人工智能、大数据、云计算为基础的支撑平台，实现高能效的智能计算系统，包括在网络端和边缘侧设备上的高效 AI 模型执行、在边缘侧的高效资源管理、在云服务器上的高效 AI 模型训练等；利用 AI 提质增效节能减排，实现面向数据驱动和 AI 决策优化算法引擎。实现面向“双碳”目标的智能绿色计算平台，利用端-边-云协同的底层的高能效 AI 计算系统，支撑上层的数据驱动的 AI 决策优化算法，比如强化学习、多智能体协同等，赋能 6G 网络绿色化。

### 5.2.8 6G OSS 基于空天地一体与通感一体的分布式自治与协同

**融合内生 AI 的分布式 OSS 技术**

6G 内生 AI 能力对应在未来网络架构的各个网元中具备 AI 能力，形成网络内分布的 AI 能力节点。各分布式 AI 节点所需的数据、网元对象、网元控制等能力依托于相应的分布式 OSS 能力提供。各分布的 AI 能力节点及相应的分布式 OSS 节点服务于所在网元所控制的功能智能化，分布式节点之间的 AI 模型共享和传输、AI 模型和训练等所需资源的调度等，可通过少量的集中式 OSS 管理节点实现管理。

**面向空天地一体及通信感知等新技术的分布式 OSS 分域治理技术**

对于超大规模天线及 RIS 智能超表面实现无线传播环境人为控制，未来网络将存在分布于无线侧的 AI 节点提供智能化计算和控制能力。对于卫星系统内的星间链路智能选择，与卫星系统星座构型和卫星服务区域密切相关，且集中于卫星系统内部，其智能化赋能也将由分布于卫星系统（如卫星载荷、卫星馈线站）。面向物联网通信智能化，未来物联网中的机器人节点不仅具备常规的通信传输能力，还可能具备对声音、气味、周围环境影像等信息的感知和识别能力，并将环境感知信息融合在物联网智能通信决策过程中。6G 网络支持的智能机器人之间的通信和网络自配置，具有明显的地域要求和数据传输质量需求，局部化的、本地化的网络智能和网络管理能力相对而言更能匹配这类高度灵活性的通信和组网管理需求。类似的，用于无线环境感知的传感器组的控制、在去小区化环境中的多收发天线对的动态配对和干扰管理等能力，采用近端部署的分布式 AI 能力节点及分布式 OSS 节点完成智能化管理。

在分布式 OSS 管理范围内，基于通感一体化的感知能力、局部网络的内生算力、网元内生 AI 能力共同构成了 6G 形态下的边、端。相较于 5G，6G 的边和端除具备更强的网络信息收集能力外，内生算力提供了边和端更多的能力空间，结合网元内生 AI 能力和分布式 OSS 管理能力，6G 的边和端对未来业务和用户需求的满足能力大幅度提升。

### 面向跨域联合的多域 OSS 协同技术

受限于分布式 OSS 管控范围，单一分布式 OSS 管理某一特定领域，如多类型传感器的环境智能感知，或某一特定空域，如特定的厂房等场景的网元控制。因此，单一分布式 OSS 管控范围内的数据类型、数据信息量、AI 训练模型等，普遍与分布式 OSS 管控对象特点密切关联。对于日常大概率反复出现的现象和控制行为，分布式 OSS 具备较好的局部管控能力。但是，对于环境改变、业务突发变动、网元突发异常等小概率事件的应对方面，分布式 OSS 可直接处理的数据及影响的网元设备相对有限，难以胜任未来业务场景中对可靠性的高水平要求。此外，分布式 OSS 管控的资源中，可用于 OSS 控制的资源相对有限，对于复杂度高的网络自治和网络资源统一管理协调任务。

针对于分布式 OSS 的局限，6G 集中 OSS 节点营运而生。集中式 OSS 节点并不是全网唯一集中节点，集中式 OSS 节点位于多个分布式 OSS 节点之间，直接负责相邻的分布式 OSS 节点之间的信息和资源的交互。集中式 OSS 节点之间传递经过汇聚的分布式 OSS 节点的信息和资源交互。

## 5.2.9 6G OSS 数字孪生网络

数字孪生网络服务作为一种新的网络服务为将在 6G 网络中提供端到端或部分网络功能的孪生服务，使能移动网络创新加速，以降低电信行业研发成本和缩短研发周期[43]。6G 网络将通过网络的数字孪生构建全新的自动化网络运维系统，实现网络全生命周期的高水平“自治”[44]。在 6G OSS 系统中基于物理网络构建数字孪生网络，结合数字孪生网络进行物理网络运维优化是 6G OSS 的关键技术。

在数字孪生网络中，各种网络管理和应用可利用数字孪生技术构建的网络虚拟孪生体，基于数据和模型对物理网络进行高效的分析、诊断、仿真和控制。数字孪生网络为网络运维优化操作和策略调整提供更接近真实网络的数字化验证环境，能在不影响网络运营的情况下完成预验证，极大地降低试错成本。通过内生 AI 和数字孪生网络的深度融合，数字孪生网络还可以预测物理网络的发展趋势和问题，基于此形成网络优化的预干预措施，并且为这些网络运维优化操作、AI 模型及工作流、智能策略提供更接近真实网络的数字化验证环境，使得验证结果更精准可靠。

6G OSS 数字孪生网络技术还包括对物理网络本体和孪生体进行编排和优化技术。根据用户的业务以及网络本身运维运营需求，统一编排所需要的资源和功能，形成所需的能力，保证用户的业务体验。6G OSS 可通过与网络孪生体交互，接受网络孪生体输出的网络配置参数，对物理网络进行编排管理，从而实现网络自动化运营，提升网络对新业务、新场景和新需求差异化需求的适应性。

## 5.2.10 6G OSS 安全内生

6G OSS 将支持 6G 网络空天地一体化立体组网，多种类型的网元设备管理编排，多种形态的网络资源连接调度。由于网络接入和连接复杂性的极大提升，

使得网络安全的边界更加模糊，传统以边界网络流量检测、分析和防护为主的“补丁式”安全措施已不能完全满足 6G OSS 的安全风险防护需求。6G OSS 系统设计时应采用安全源于设计（Security By Design）的原则，保障 OSS 自身以及所提供服务和决策的原生可信，通过提供基于流程而非边界的安全防护技术，实现在分布式异构资源编排调度、自治管理以及能力开放过程中的潜在安全风险防范，使 OSS 具备应对不确定安全威胁的能力，从 OSS 系统内部提供内生安全的全面保障。

### 6G OSS 面向异构资源的信任融合技术

6G OSS 需确保端到端资源编排的可信，基于区块链技术在不同网络实体之间构建融合信任，通过设置动态安全监控以及管理控制，实现分布式 6G OSS 网络资源调度编排，信令传输执行反馈的全业务流程的安全可控。通过提供持续自适应风险和信任评估，在资源编排交互过程中进行风险管理，不断地监视和评估风险信任级别，如果发现信任下降或者风险增加到达阈值，需要进行响应，及时地调整网络资源的连接调用策略。

### 6G OSS 基于 AI 的内生安全能力编排技术

6G OSS 应提供可供调用的安全能力资源池，基于分布式 6G OSS 多元本地环境以及集中式 OSS 协同生成安全策略。通过智能编排的方式与其他网络设施或服务一起形成柔性的按需服务，提供基于 AI 智能生成的主动安全防护策略，能够同步性甚至前瞻性地适应网络变化，完成安全资源编排，以衍生网络内在稳健的防御力，实现安全能力弹性部署，提升网络韧性。此外，6G OSS 的内生安全技术应具备更小的计算开销、能量开销能力，以满足网络低功耗、绿色的需求。

### 6G OSS 泛在协同安全态势感知技术

6G OSS 支持基于联邦学习、隐私计算等技术，通过大数据分析、异常检测、态势感知、机器学习等技术，实现态势感知端、边、网、云的智能协同，准确感知整个网络的安全态势，敏捷处置安全风险。提供基于 AI 和大数据分析的实时风险分析，及时预警响应如窃听、干扰、节点假冒、伪造、篡改、数据泄露等潜在的安全风险。支持基于围绕资源和业务不同等级的安全需求，融合安全和网络管理，实现风险告警到安全联动响应的自动闭环。

### 5.2.11 6G OSS 算力内生网络管理

，算力内生网络突破传统网络服务和算力服务的实体边界，在通信网元功能基础上，利用通算智能调度编排决策机制，实现新型算力业务的灵活接入与即取即用。算力内生网络/专网产品的总体架构主要由三部分组成，算力内生无线基站：提供通信网络无线接入能力并通过网络负载空闲算力为边缘应用及网络 AI 提供算力服务；算力内生 MEC：实现边缘计算平台 MEC 功能并提供通算资源一体调度与编排功能；算力内生核心网网元：实现轻量化核心网功能，支持能力开放与 MEC 的通算协调调度[45]。

6G OSS 的算力内生网络管理技术可从通算网元动态决策、网络应用无损迁移，云网算力协作三个方面进行研究。

**通算资源动态决策**

6G OSS 作为算力内生网络的通算调度与编排决策器，将边缘应用部署在算力内生网络/专网中，支持边缘算力应用的无损迁移及算力均衡。在无线网络业务闲时，通算调度与编排决策器将通算网元的空余算力抽象成动态算力节点，进行纳管，提供给边缘应用作为算力资源，避免了闲时无线基站的资源的浪费，提高了无线网络设备的利用效率。反之，当通信网络业务忙时，通算网元算力资源紧张，OSS 则将应用进行收缩和迁移，迁移应用至固定算力节点或者空闲算力节点，以支撑此时资源紧张的无线业务，从而达到无线算力资源动态管理目的。同时，通过对全网内生算力资源的实时采集分析，6G OSS 将应用基于联邦学习的网络级内生算力调度算法，均衡全网内生 MEC 算力资源池的算力资源利用率。

**网络应用无损迁移**

6G OSS 将提供网络应用调度和迁移服务管理能力，通过将 MEC 上的边缘网络应用分解用最小能力集和扩展能力集，优先将最小能力集部署在固定算力节点，然后灵活地将扩展能力集延伸到动态算力节点。在无线算力资源紧张的情况下，可以只迁移扩展能力集，这种方式能最大保证边缘应用的迁移无损，还能有效提高无线资源的利用率。

**云网算力协作**

6G OSS 还将具备云算力与网络内生算力的协作技术，根据算力业务类型和业务应用场景，编排调用云网算力资源，满足客户网内和网外的算力业务需求，实现云网算力的深度融合。

### 5.2.12 6G OSS 智慧内生

面向 6G 网络智慧内生的需求，6G OSS 将提供支持基于内生 AI 的端到端网络服务和资源的编排、管理和调度。此外，6G OSS 通过搭建统一的 AI 信令体系，支持适用于 6G 分布式网络架构以及与具备内生 AI 的网元协同，支持 AI 服务与多类网络资源服务间的信息交互，实现 AI 能力与其他网络服务无缝融合。6G OSS 还将提供统一的 AI 全生命周期管理，实现分布式网络的 AI 内生与网络功能的一体化编排。

#### 6G OSS AI 能力组件化动态编排技术

6G OSS 需要面向端到端的云原生网络，以服务化的方式为网络提供各类 AI 能力和工具，包含 AI 服务的动态发现、组合和编排。应提供 AI 服务组件化的管理与编排，通过定义通用和多元化 AI 服务，支持基于意图分析，动态匹配的 AI 能力。支持联合多类网络资源和 AI 服务需求，联合进行资源的分配、编排，实现 AI 与网络资源服务的灵活解耦或组合。协同不同网络节点作为 AI 训练/执行的一部分，根据本地收集的数据集，通过大量相关联的设备共同生成符合本地区域或场景的 AI 服务编排模型。

#### 6G OSS 分布式内生 AI 协同管理调度技术

6G OSS 将提供分布式、跨域 AI 能力的统筹协同管理调度，6G OSS 通过提供 AI 模型的全生命周期工作流，支持基于特定服务需求调度全网 AI 能力，实现全网 AI 能力合理分布。6G OSS 将支持基于 AI 的分布式多层级的网络资源编排，建立分布式实时协作模型生成、训练、推理框架，实现多维度网络、存储、计算等异构资源的融合编排。6G OSS 自身的 AI 能力也将以分布式与集中式相结合的形式，因此需支持部署在不同网络位置的 AI 服务之间的通信、资源分配和编排。通过提供 AI 模型的局部或全局的可解释性方法或接口，实现 AI 模型导入和重用的机制与接口，提供分布式模型训练框架，支持基于上下文的自动化调参技术。

### 6G OSS 基于数字孪生的 AI 模型仿真验证

6G OSS 应提供数字孪生技术支持的生成测试数据和测试场景的仿真环境，提供导入 AI 模型验证指标及测试用例的通用接口，提供从模型验证结果反馈到 AI 工作流各环节的闭环优化流程，提供对外接口对模型进行形式验证。建立数据与模型的监控技术体系，确保 AI 模型、AI 服务的多维风险感知，对模型推理结果的应用效果进行预判，提供持续的模型在线更新机制及相关接口。

### 6G OSS AI 信令交互体系

分布在网络不同位置的 AI 能力需要在 6G OSS 系统中通过一套标准的 AI 信令体系实现 AI 能力的交互协作。AI 信令交互流程，包括需求发起过程各 AI 网元间的信息交互和模型实现过程中协作运算，并需以鉴权为切入点，以资源服务为基础，最终实现 6G 网络灵活注智。

### 6G OSS 多维内生 AI 评估评价体系

6G OSS 需基于统一的 AI 服务的评估评价体系，管理 OSS 内生 AI 服务。构建基于环境上下文动态优化的、综合性能与效率的模型评估，评价维度应涵盖 AI 模型以及 AI 服务的设计、建模、复杂度、性能能效、调用频率以及服务反馈等多方面因素。针对 AI 模型的鲁棒性、安全性、预测或识别的准确性建立环境相似性的模型度量机制以支持迁移学习、联邦学习。针对分布式内生 AI 能力，需提出针对特定场景的 AI 模型同步评估指标，如传输要求，AI 性能要求等。提供用于 AI 伦理、偏见、风险控制以及识别 AI 模型本身无法做出正确的决策或做出错误决策造成的重大损失的关键问题要素或风险清单。

## 5.3 6G OSS 技术框架小结

通过对以上 12 项潜在关键技术的研究，我们从 OSS 的内生支撑能力和网络管理能力两个层面构建了 6G OSS 的技术框架，如图 5-5 所示。

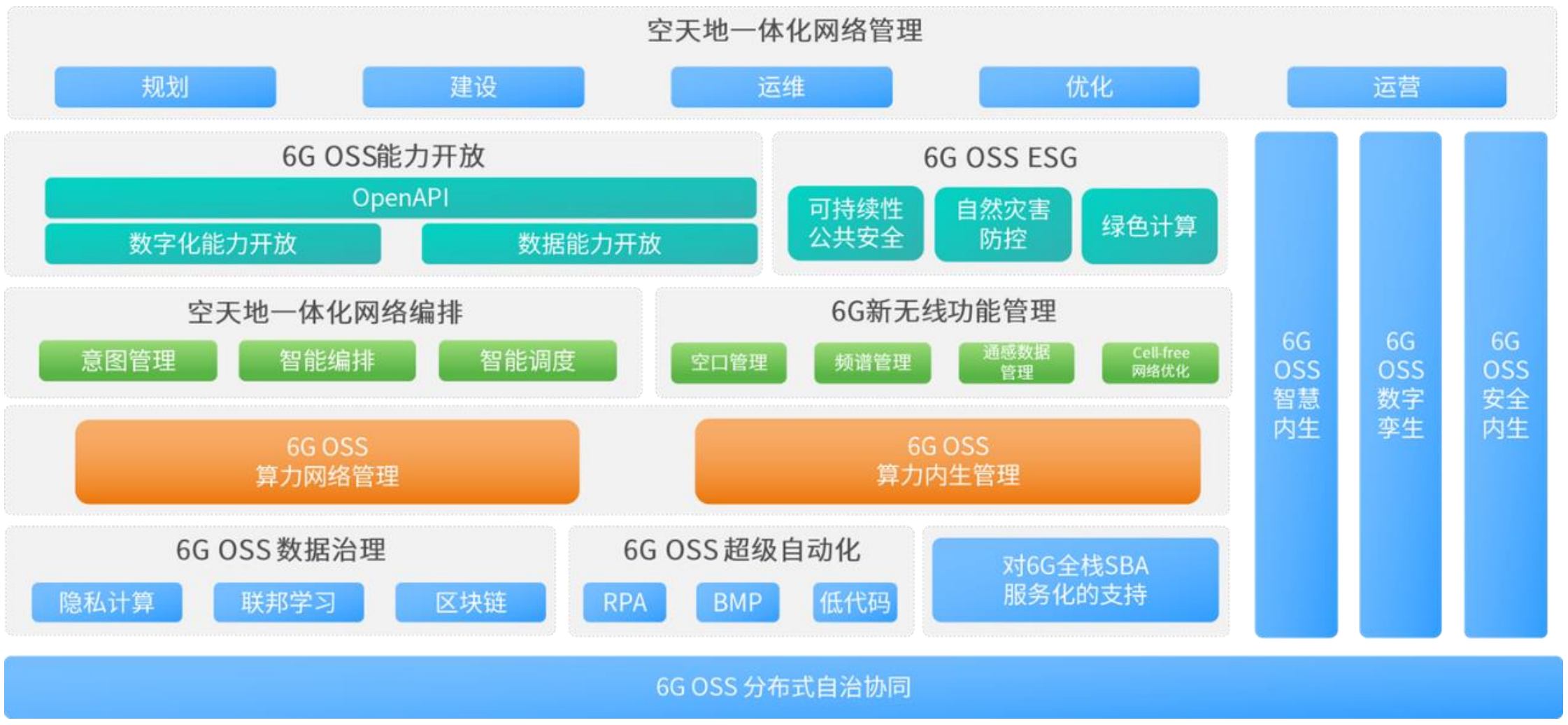

**图 5-5 6G OSS 技术框架**

# 六. 6G OSS 架构与功能

## 6.1 6G OSS 功能架构

为了满足 6G 关键业务场景和网络技术演进的需求，6G OSS 系统的功能架构（如图 6-1 所示）主要由 6G OSS 核心功能、三大通用能力和三项管理功能构成。

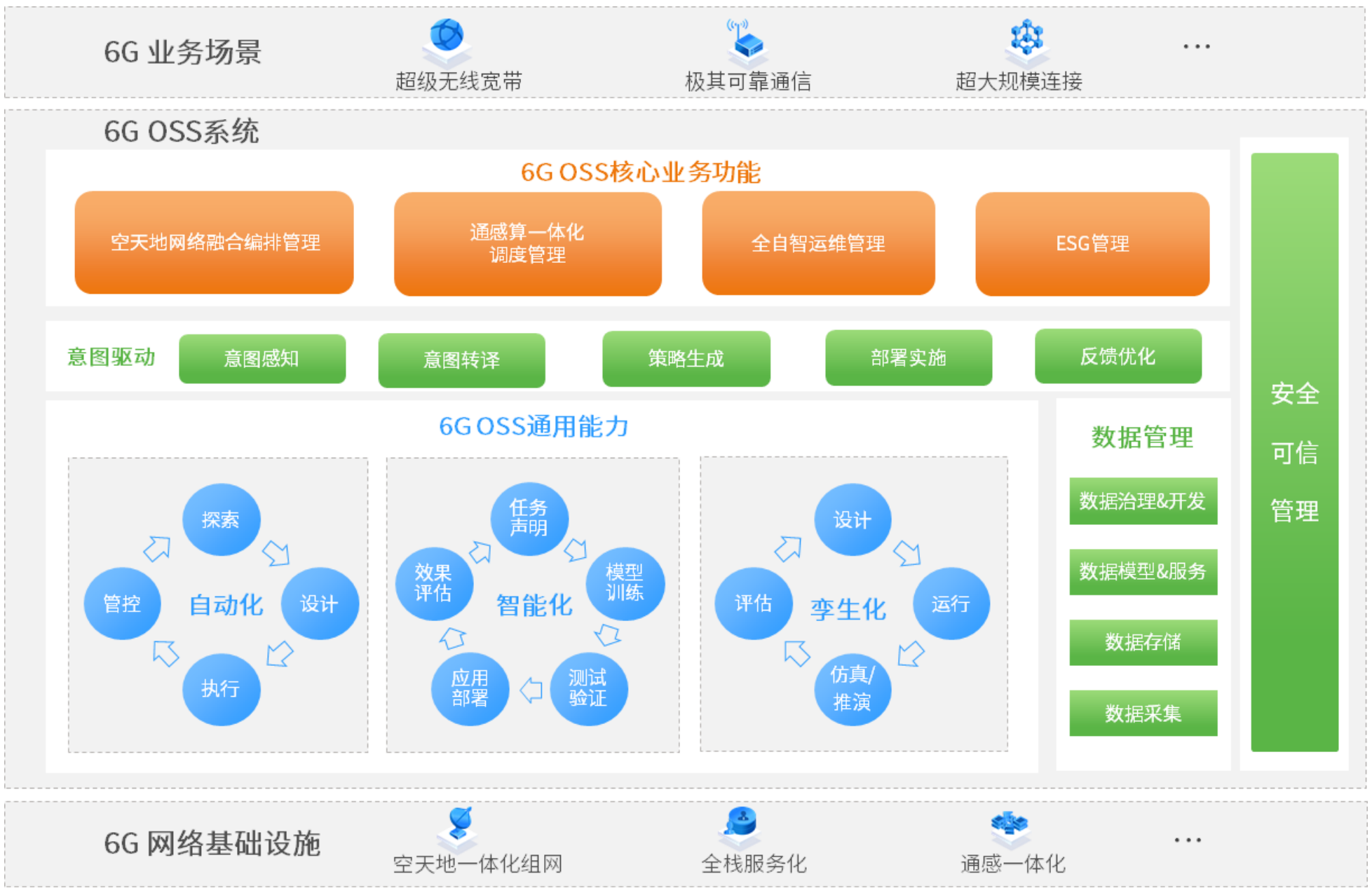

图 6-1 6G OSS 功能架构

- **6G OSS 核心业务功能**：负责网络“规、建、优、维、营”全生命周期中的网络管理业务功能，具体包括空天地网络融合编排管理、通感算一体化调度管理、全自智运维管理和 ESG 管理等功能；
- **意图驱动功能**：负责衔接 OSS 业务功能产生的业务意图与核心能力闭环，具体包括意图感知、意图转译、策略生成、部署实施、反馈优化；
- **自动化能力闭环**：负责网络管理能力的执行功能，通过探索、设计、执行与管控的自动化闭环，赋能 6G OSS 功能实现自动化。
- **智能化能力闭环**：负责网络内生 AI 模型的全生命命应用运行提供所需的管理能力，通过任务声明、模型训练、测试验证，网络应用和效果评估五个 AI 开发应用步骤的闭环，赋能 6G OSS 功能实现智能化。

- **孪生化能力闭环**：负责 OSS 系统的网络孪生化，通过构建物理网络的数字孪生体，实现基于孪生化的网络全生命周期的分析、诊断、仿真和控制，通过预测性运维实现网络的高度自治和全面的网络智能化。
- **数据管理功能**：负责 6G 网络运力、算力与存力等相关资源、告警、性能、质量等数据的采集、存储、计算、治理、建模与服务等。
- **安全可信管理功能**：负责 OSS 系统安全和可信领域的管理，通过信任管理、访问权限管理、安全策略管理、安全态势感知、弹性容灾管理实现 OSS 全流程的安全可信。

## 6.2 6G OSS 功能描述

### 6.2.1 6G OSS 核心业务功能

6G OSS 核心业务功能直接承载运营商 6G 网络运维管理需求，6G OSS 不仅要囊括 5G OSS 系统中典型的网络资源管理、故障监控、网络性能与质量管理、业务编排等能力，还应面向 6G 业务场景和需求，基于意图驱动和自动化、智能化、数字孪生化三闭环的支撑能力，从核心功能上进一步演进升级，为 6G 空天地一体化网络的“规、建、优、维、营”全生命周期提供运维保障能力，支撑 6G 全业务场景。6G OSS 的潜在业务功能建议包括：空天地网络融合编排管理，通感算一体化调度管理，全自智运维管理和 ESG 管理等。

空天地网络融合编排管理实现网络业务编排，空天地网络融合编排和通感算资源弹性编排能力。网络业务编排面向业务需求，通过意图网络驱动，使用网络抽象语言将各种网络功能和服务单元进行有序的组织和拉通，形成可自动化部署的网络功能集合，提供可动态调整、重复使用、快速创新的网络服务能力。空天地网络融合编排实现各域网络之间的网络功能、网络节点、网络资源、网络拓扑和网络频谱的协同组织和拉通，基于自动化和孪生化闭环提供高可靠、高灵活性和高可拓展性的网络服务能力，适应动态多样的空天地业务场景需求。通感算资源弹性编排实现通信、感知、计算等资源的编排管理，在满足业务 SLA 的前提下，基于性能、资源、经济、绿色、安全等多因子的综合决策，通过智能化闭环实现弹性和最优的通感算资源编排，保障用户业务 SLA 和资源的联合最优。空天地网

络融合编排管理北向对接业务需求，南向对接通感算一体化调度管理并为其提供编排调度方案。

通感算一体化调度管理承接空天地网络融合编排管理的调度要求，是基于编排方案的资源调度管理执行中心。基于自动化闭环执行 6G 网络基础设施的通感算资源自动化调度，包括空天地、通感算资源节点的拓扑管理、监控管理、策略管理、配置管理、性能管理、质量管理、数据管理等功能。同时，通感算一体化调度管理提供分布式、去中心化、自治化的组网能力，实现资源的动态共享和立体部署，支撑分布式自治的 6G 网络体系架构。通感算一体化调度管理北向对接空天地网络融合编排管理，南向对接通信、感知、计算等 6G 新的专业管理单元，如新无线网络管理、新核心网管理、新传输网管理等。

全自智运维管理提供 6G“规、建、优、维、营”全生命周期的端到端自动化、智能化、孪生化网络运维能力。目前自智网络的 L5 级自动化、智能化目标将是 6G 全自智运维管理的基础能力和第一阶段，在此基础上，6G 全自智运维管理将进一步扩展自智网络的内涵和外延，实现针对通感算资源、跨空天地网络子域和 B/O/M（业务、运维、管理）全域的自动化、智能化、孪生化运维管理。现有 5G OSS 中的资源管理、网络故障管理、性能管理、质量管理、运维管理等能力将纳入 6G 全自智运维管理中，成为其一体化闭环管理的环节组成。6G OSS 全自智运维管理基于通感算资源的综合管理和跨网络子域和 B/O/M 多域融合的数据共享，并通过内生 AI 和数字孪生网络，自动实现多业务、多领域、全生命周期、全场景的闭环运维管理，满足 6G 网络自智（AI4NET）和网络普惠智能（NET4AI）的目标愿景。

ESG 管理是赋能 6G 可持续和高质量发展的核心能力，也是 6G 满足社会服务均衡化、高端化，社会治理科学化、精准化，社会发展绿色化、节能化目标愿景的关键支撑能力。ESG 管理基于空天地一体化网络提供可持续的公共安全网络保障与管理能力；基于空天地一体化组网和通感一体化提供全域覆盖、感知增强的自然灾害防控能力，结合数字孪生，实现公共安全、自然灾害等灾难的预警、灾难模拟推演及灾后恢复。同时，ESG 管理还将基于高效能绿色计算技术降低 6G

系统本身的能耗，打造绿色数字基础设施，并进一步支撑绿色普惠的数字化创新业务生态，助力“双碳”目标。

### 6.2.2 意图驱动功能

6G OSS 的意图驱动功能包括意图感知、意图转译、策略生成、部署实施、闭环优化等具体功能。

- **意图感知**：提供全面感知业务意图的能力。面向未来 6G 空天地一体化和通感一体网络形态、沉浸式全息通信业务需求以及多元化交互，意图输入节点和方式将呈现分布式、多元化、立体化特点。支持通过指令、文字、语音、动作、触觉等多种交互方式以及不同协议报文的自动化采集意图。基于自然语言处理、语义分析，结构化输出网络意图，助力 6G OSS 由面向网络节点连接的开通运维演进为面向多维资源匹配的融合协同。
- **意图转译**：将业务意图解析、转换为网络能力。基于业务场景，结合人工智能、网络遥测等技术将业务意图解析为如资源、性能、路径等网络能力需求。在此基础上，基于知识图谱技术融合网络领域知识，实现自动化网络需求预测，智能化网络能力匹配，将网络能力需求转换为可配置执行的机器可读语言，并确保意图在解析转译过程中的完整性、去模糊性。
- **策略生成**：根据网络能力要求提供网络开通、组网方案等策略生成能力。提供意图驱动的可编程、可拓展的端到端动态组网策略以及编排调度方案，实现灵活的业务定制开通，智能化网络管理及运维。结合网络资源效率和能力分布，动态生成匹配业务意图以及网络能力的策略。面向动态场景需求和演变趋势，基于机器学习以及增强学习，提供灵活的动态寻优策略。
- **部署实施**：提供组网方案实施以及动态编排调度能力，主要包括意图验证、意图冲突解决、网络资源编排配置。意图验证提供基于数字孪生技术的虚拟验证环境，验证基于意图的网络策略是否按预期执行并能够达到预期效果。意图冲突解决可实现动态意图冲突识别并自动化生成协调策略。网络资源编排配置基于智能化策略匹配、跨域意图编织技术以及数字孪生技术，实现无需人为干预的策略执行的自动下发、运行监测以及反馈闭环。
- **反馈优化**：提供网络性能感知、需求能力匹配度监控及优化能力。基于配置、监控、分析、管理、评估、优化全面提升意图驱动的网络全生命周期的运行效率以及资源使用效率，保障业务意图如愿达成。基于多目标优化

技术以及动态优化策略，沉淀典型场景下的策略模板，全面提升网络内生 AI 的自主柔性，实现意图网络的全面反馈优化。

### 6.2.3 自动化能力闭环

如图 6-2 所示，自动化闭环作为 6G OSS 通用能力闭环之一，主要基于全面的 6G 数据管理，通过对 6G OSS 业务功能相关任务与流程的端到端探索、设计、执行与管控管理，赋能 6G OSS 业务功能实现任务与流程的自动化。

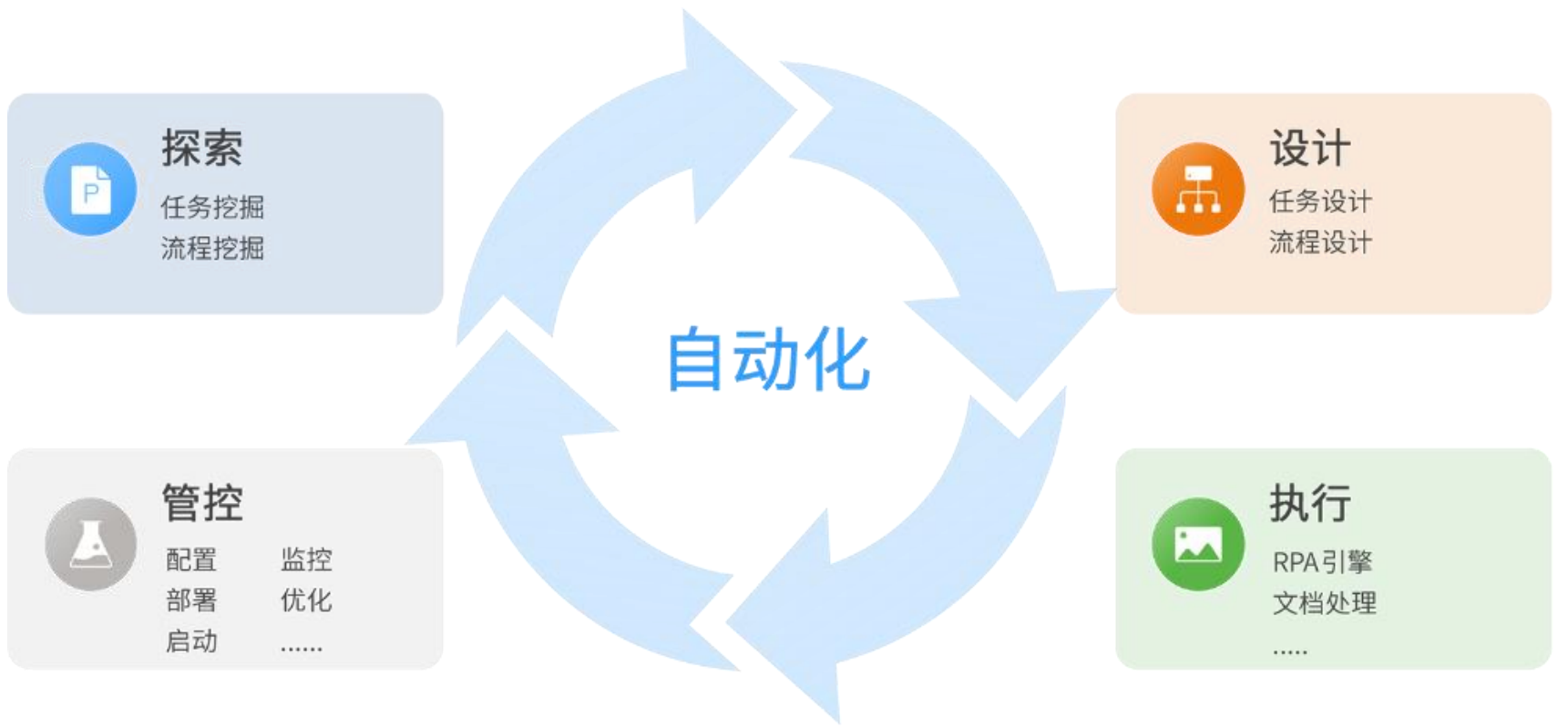

**图 6-2  6G OSS 自动化能力闭环**

相比于自智网络“感知，分析、决策与执行”的 4 环节闭环，6G OSS 基于超级自动化技术，结合 RPA、流程挖掘与低代码/零代码等技术，对 6G OSS 业务功能的任务与流程的自动化给出了具体的方法与实施路径，助力运营商 6G OSS 快速有效地实现其业务功能的自动化。

- **探索**：6G OSS 基于其超级自动化的流程挖掘（Process Mining）/业务流程管理(BPM)等技术，实现 6G OSS 业务功能的任务挖掘与流程挖掘。其中任务挖掘主要实现自动识别并汇总 6G OSS 业务功能相关的操作记录，识别具有高度自动化潜力的步骤，如对空天地一体化网络编排管理功能中的业务开通操作进行挖掘，实现业务开通过程中资源勘查、预占与配置等操作的自动化。流程挖掘主要使用 6G OSS 业务应用（如空天地一体化网络编排管理、协同调度管理与全自智运维管理等）中留下的数字足迹，进行流程还原，并自动找出流程痛点和瓶颈，并通过自动化闭环中的设计等环节进行改进与优化。
- **设计**：6G OSS 基于其超级自动化的低代码与零代码等技术，通过一个简单的拖放式编辑器，快速设计自动化流程。针对 6G OSS 业务功能的工

作流，如空天地一体化网络编排与智能运维工作流等，只需通过屏幕录制记录工作流，无需进行手工编程，即可完成工作流程与任务的设计构建。

- **执行**：6G OSS 基于其超级自动化的 RPA 等技术，实现设计环节的完成的 6G OSS 工作流与任务的自动化操作与执行等。其中机器人引擎支持有人值守、无人值守和高密度三种形式，实现 6G OSS 主要功能流程的高效、稳定运行，如空天地一体化网络编排资源勘查流程自动执行等。另外，6G OSS 通过智能文档处理功能，支持利用预先训练好的 AI 算法模型，进行非结构化数据（如表格、文档、图片、音视频）的自动识别、分类、要素提取、校验、比对、纠错，如在全自智运维管理中的机房或网络自动巡检，可实现设备与机柜等图像自动识别，及时发现异常并执行后续的网络自智运维等。
- **管控**：提供配置、部署、启动、监控、测量和跟踪 6G OSS 自动化工作流或任务所需的能力，确保全部自动化流程与任务的安全高效执行。

对于 6G OSS 工作流程设计的任务拆解，建议仍参考 TMF 遵从以下三个原则：

- 完整性：所有操作维护动作必须要能拆解到上述的五个步骤中。
- 平衡性：各个任务的大小和粒度必须基本一致。
- 互不重叠性：各个任务需为原子粒度，承载的功能互不交叠。

### 6.2.4 智能化能力闭环

智能化闭环是 6G OSS 通用能力闭环之一，主要为 6G 网络内生 AI 模型的全生命应用运行提供所需的管理能力，通过任务声明、模型训练、测试验证，网络应用和效果评估五个 AI 开发应用步骤的闭环，赋能 6G OSS 功能实现智能化。图 6-3 展示了智能化能力闭环的工作流程。

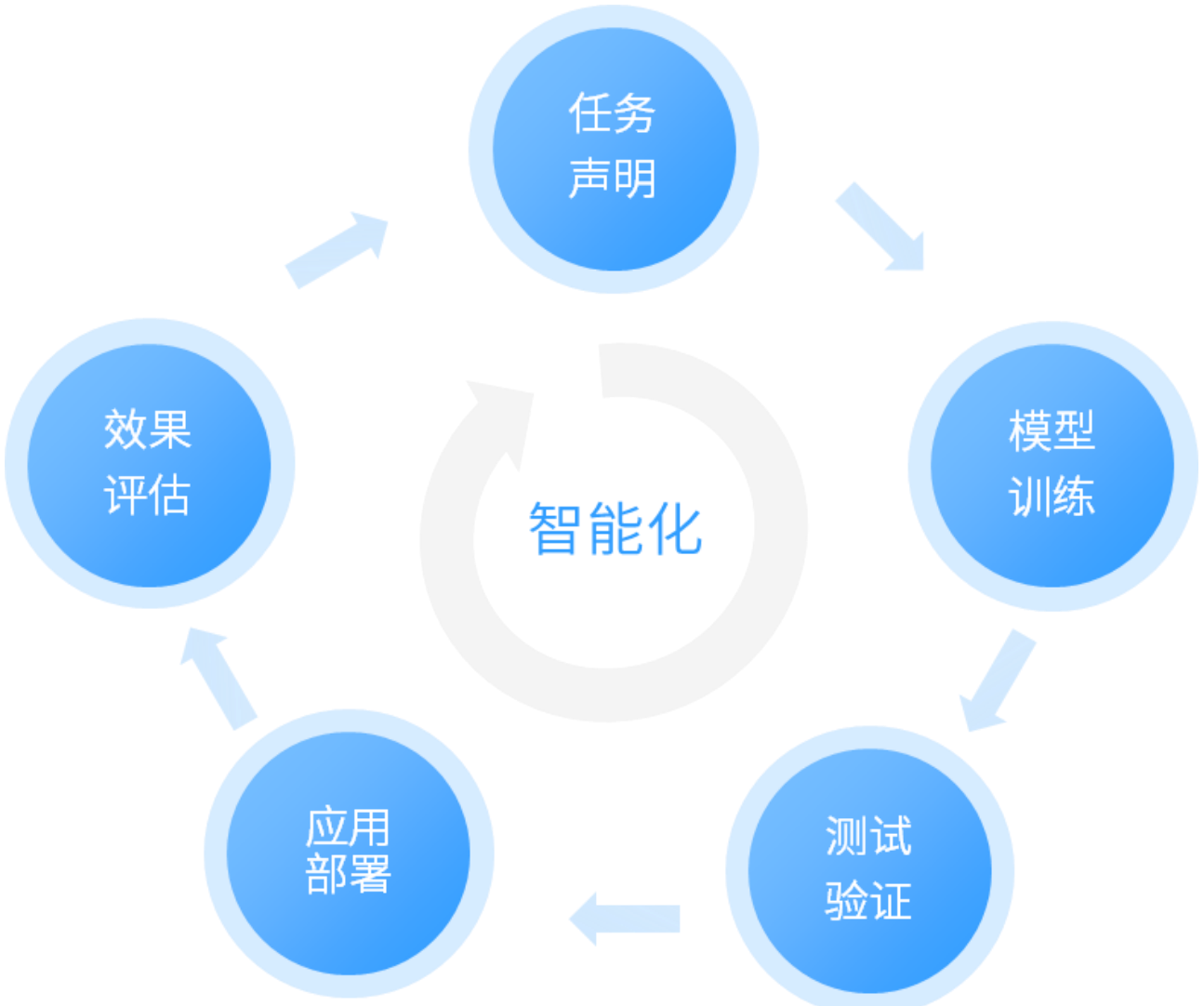

**图 6-3 6G OSS 智能化能力闭环**

- **任务声明**：网络 AI 任务声明是内生 AI 的全生命周期的起始步骤，6G OSS 系统可根据通过与意图管理套件的交互，获得网络业务意图，并根据业务意图生成对于 AI 任务的描述，包括任务目标、任务场景、任务所需的 AI 模型类型等。
- **模型训练**：根据网络 AI 场景的描述，OSS 系统选择调用相应的算法并应用对应的场景数据中的训练数据进行模型训练，生成 AI 模型。
- **测试验证**：OSS 系统应用场景数据中的测试数据对于 AI 模型进行测试验证，对于无监督模型等无法进行数据测试的模型，可通过基于业务逻辑的设定规则进行模型的验证。
- **应用部署**：对于通过测试验证的模型，OSS 系统需要负责 AI 模型与相关网络功能的应用对接，通过定义 AI 模型的输入输出数据格式、参数配置等，实现网络功能对于模型的自动调用。
- **效果评估**：对于已经在网络中部署应用的 AI 模型，OSS 系统需要对模型进行周期性或事件性的效果监控，评估 AI 模型的运行效果是否满足场景设计需求并判断是否需要进行模型优化。

## 6.2.5 孪生化能力闭环

孪生网络由一系列孪生体构成，而且孪生网络提供了物理网络的各种仿真操作。通过仿真任务完成仿真场景的编排，按需选择合适的网元和模型后，通过仿真任务实例化触发孪生体进入运行态，得到仿真结果。孪生体由外观模型、若干

功能模型和性能模型组成，孪生体进入运行态，意味着功能模型和性能模型被实例化。网络拓扑是孪生网络全息可视化的骨架，根据采集到的配置数据生成，物理网络同步上来的实时数据叠加到网络拓扑进行呈现，最终实现物理网络的可视可管。如图 6-4 所示，孪生化能力闭环由设计、运行、仿真/推演、评估四个步骤组成。

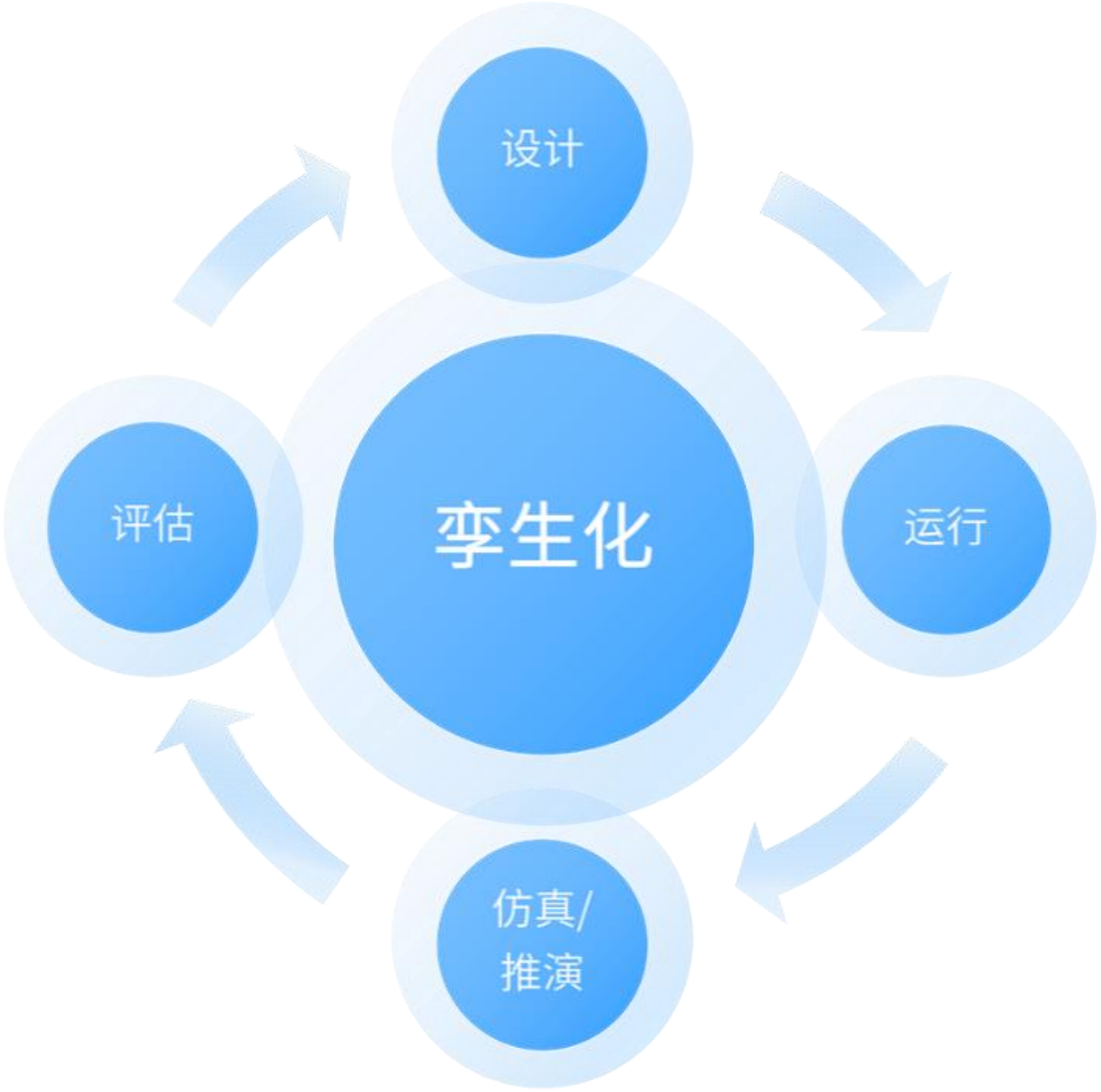

**图 6-4 6G OSS 孪生化能力闭环**

### 数字孪生网络设计

数字孪生网络设计包括数字孪生单体建模和数字孪生应用场景建模。设计阶段也称为数字孪生网络设计态，主要完成物理实体单体和应用场景的孪生设计定义。单体建模基于本体理论实现实体的表征，首先定义本体的组成要素，例如，类、属性、关系、规则和实例等多元组元素，继而通过本体模型对大规模网络数据进行一致性表征。面向通信网络设备、逻辑网元，根据物理设备信息、环境信息、拓扑节点信息、网络链路信息、容器虚拟机信息、网元配置信息等建立无线网络数字孪生体单体模型，包含物理空间模型、机理模型、语义模型等，实现数字空间和物理实体的关联，最终实现对真实网络的实时精确建模。在“规划、建设、运维、优化”网络全生命周期中，为数字孪生的仿真、可视化及智能运维能力提供基础能力支撑。

数字孪生应用场景建模包括数据模型建模和网络能力建模。数据模型建模基于网络运行数据及各类指标，实现对网络各类特征模型的智能构建。网络能力建模根据网络运行数据及指标，实现数字孪生体的高保真模拟，同时支持在场景构建中实现业务规则的验证测试。

### 数字孪生网络运行

数字孪生网络运行是指基于数字孪生网络设计得到的模型和场景，根据输入孪生模型的各项随时间和空间动态变化的数据，通过数字孪生的数据模型和网络能力模型模拟得到通信网络的性能表现的过程。运行中的数字孪生网络通常称为数字孪生网络的运行态，此时数字孪生网络实时同步物理网络的运行情况。在数字孪生网络运行态下，输入的数据根据具体的应用场景不同，可以是网络运行的历史数据，也可以是基于网络推演或预测的网络数据，或历史数据和预测数据同时输入。运行的输出结果是根据数字孪生网络场景设定的数字孪生网络孪生体输出指标。

### 数字孪生网络仿真推演

数字孪生网络仿真推演过程也称为数字孪生网络处于仿真态。数字孪生网络仿真态基于某个运行态的时间切片，形成多个物理镜像，进行模拟推衍。其方法分为模拟网元属性和网络通信协议的网络性能仿真方法，和基于人工智能的网络性能模拟方法。前者通过对网络通信协议的规则的计算机模拟，推导在一定的网络输入数据条件下，网络性能表现，移动通信传统网络性能仿真即采用该方法。由于仿真精度直接受到对网络通信协议建模精细程度的影响，在仿真所需计算量、网络模型随通信协议标准更新带来的代码更新方面耗费大量人力物力。应对基于网络通信协议的网络性能仿真不足，基于人工智能的网络性能模拟方法迅速崛起。基于人工智能的网络性能模拟，将网络输入数据和网络性能指标分别作为模型的输入和输出，通过人工智能算法生成网络行为的人工智能模型。通过将网络数据代入训练好的网络模型推演网络性能指标，从而减少编写精确的通信网络协议代码的人工消耗。为进一步提升基于人工智能的网络性能模拟结果精度，可代入通信协议的规则模型形成知识图谱，降低人工智能模型设计和训练所需的数据和运算量。

**数字孪生网络评估**

数字孪生网络评估是基于数字孪生网络对不同输入数据条件下的网络性能结果进行对比分析的过程。评估方法可基于收集的大量专家经验形成的评估规则、通过网络模拟得到的人工智能模型，或两者结合的方法。在大型网络中，基于专家经验数据训练的人工智能模型可有效的减少网络运维人工投入的同时，保持较好的网络自治性能。

## 6.2.6 6G OSS 数据管理功能

6G OSS 数据管理主要实现 6G 网络运力、算力与存力等相关资源、告警、性能、质量等数据的采集、存储、计算、治理、建模与服务等，提升 6G 数据研发效率、降低数据管理成本、赋能 6G 数据价值流通。

如图 6-5 所示，6G OSS 数据管理主要包括数据采集接入层，数据存储计算层，数据管理层与数据模型&服务层。

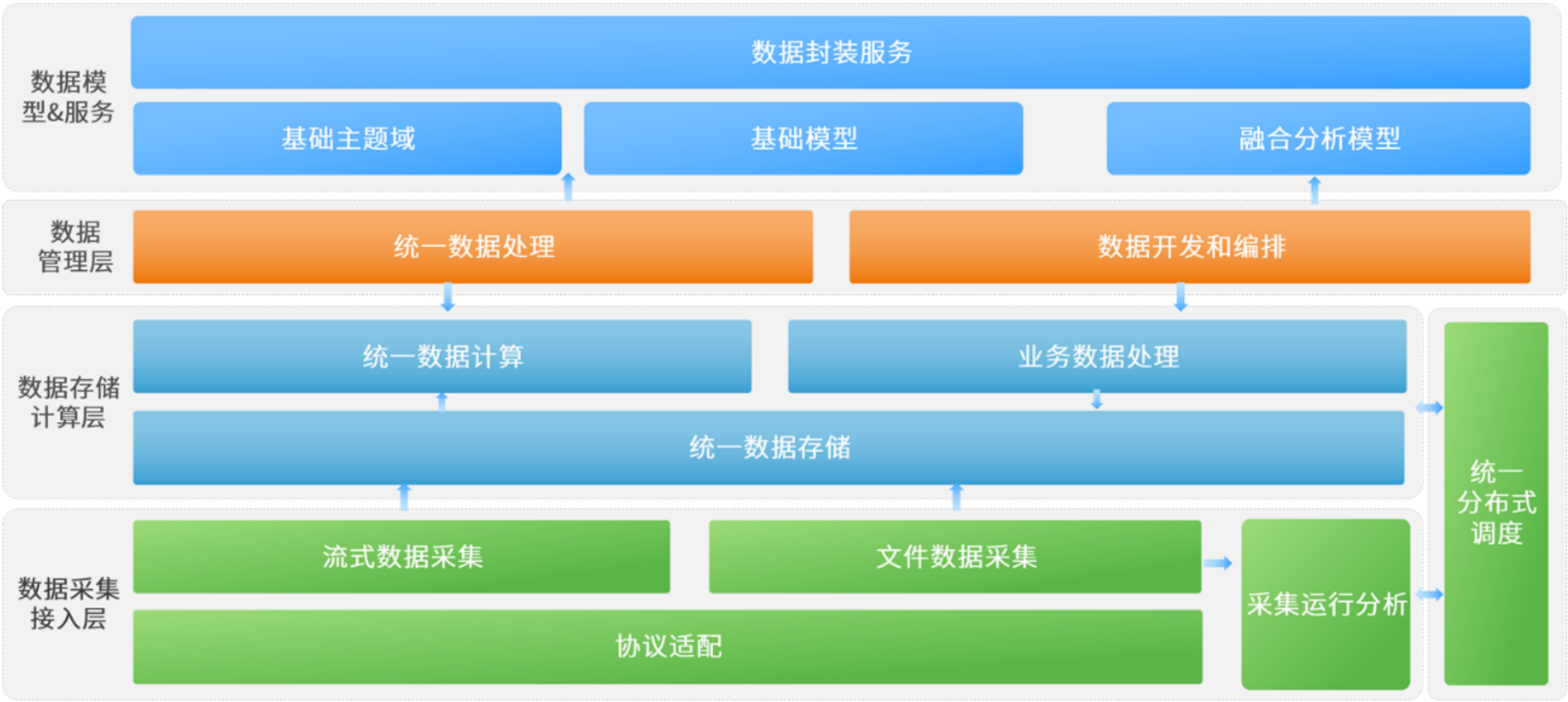

**图 6-5 6G OSS 数据管理**

- **数据采集接入层**：实现 6G 网络、计算、存储相关资源、告警、性能与质量等数据及 NWDAF 等网络功能即时分析数据的采集。数据采集接入层主要包括协议适配，流式数据采集，文件数据采集与采集运行分析。其中协议适配主要实现 6G 多维数据采集统一的协议适配与管理，如 Restful、SNMP 等；流式数据采集主要实现 6G 网络实时流式数据的采集；文件数据采集主要实现文件数据解压缩、文件分割合并、格式转换、数据补采策略与采集数据质量检查等。同时，通过采集运行分析功能，可实现采集数

据业务数据量分析与采集接口机/采集集群状态分析功能，保证 6G OSS 数据采集质量。

- **数据存储计算层**：实现 6G 网络、计算、存储等多维数据的存储与数据处理。其中统一数据存储主要包括分布式文件存储、分布式 K-V 存储、分布式数据仓库、内存数据库、分布式关系型数据库与分布式多维索引数据库等。统一数据计算主要实现数据批量计算，实现数据实时计算。业务数据处理实现对数据的清洗转换，关联回填，多维分析与挖掘和预测等。同时数据采集层与存储计算层的功能，可通过统一分布式调度功能模块对外提供数据采集与存储计算能力。
- **数据管理层**：主要实现 6G 多维数据的统一数据治理，数据开发与编排。其中统一数据治理基于机器学习和自然语言处理技术，实现智能元数据管理，快速整理高频词根并将数据标准与元数据自动映射，建立数据标准管理体系；通过联邦学习提升数据治理能力，实现高效数据流通管理；通过机器学习自动识别数据质量，对数据质量进行效果评估和智能修复，提升数据管理能力。数据开发和编排实现数据模型的开发和数据开发或应用流程的编排，赋能上层数据模型服务层。
- **数据模型&服务层**：主要基于数据管理层实现 6G 网络数据的基础主题数据域模型、基础模型、融合分析模型的管理，并提供数据封装服务。其中基础主题数据域管理主要实现 6G 网络、计算与存储等数据的资源域(如网络资源、计算资源、存储资源与服务资源等)，质量分析域（如各项资源与业务运行的性能指标情况等）、事件域（如告警、性能异常等）、配置域（如局数据、设备参数等）、运维域（如工单、日志、作业等）与知识域（如专家知识、方案与规则等）等数据主题域管理。基础模型管理主要实现 6G 网络资源模型、性能模型、无线网优模型、详单数据模型、通感一体数据模型、集客家宽模型与故障模型等数据模型的管理。融合分析模型管理主要实现 6G 网络融合质量分析、资源分析与业务感知分析等模型的管理。同时，通过数据封装服务将各类数据模型能力进行组合封装，对外提供数据服务，如实时查询、多维性能分析、质量洞察与数据订阅服务等。

### 6.2.7 安全可信管理功能

6G OSS 的安全可信管理包括信任管理、访问权限管理、安全策略管理、安全态势感知和弹性容灾管理等具体功能。

- **信任管理**：主要包括身份验证、密码管理、隐私管理、可信认证等功能。面向 6G 网络“规、建、维、优、营”的业务需求，通过区块链、隐私计算、联邦学习等技术，实现分布式网络异构节点、资源、能力以及服务的接入、管理、调用、开放等全生命周期的信任管理，保障 OSS 编排调度，信令传输、执行反馈的全业务流程的安全，降低隐私泄露风险。
- **访问权限管理**：主要包含权限管理、访问控制等功能。权限管理可实现 6G OSS 系统基于业务场景、用户角色以及行为的自动化灵活配置账号的数据权限、操作权限和页面权限。访问控制主要实现对通感算多维资源以及数据的访问管理，支持基于身份、行为、属性、角色、规则等多维度组合管理，实现动态细粒度访问控制，保障最小权限访问。
- **安全策略管理**：主要包含风险评估、安全审计、合规分析、策略优化等功能，匹配业务需求、资源现状、网络优化需求以及安全保障需求，通过持续自适应评估分析，自动生成动态安全策略，为管理者提供安全风险评估和应急响应的决策支撑，保障 6G OSS 信息安全的机密性(confidentiality)、完整性(integrity)、可用性(availability)。
- **安全态势感知**：主要包含态势感知、风险识别、监测预警、响应处置等功能。基于网络监测数据以及业务分析数据，提供数字孪生技术支持的通感算多维资源和空天地一体化网络安全态势可视化交互展示，基于 AI 算法以及异常检测模型，自动识别潜在安全隐患和漏洞，实现网络运行维护的安全态势感知、风险扫描、应急响应的智能化闭环处置。
- **弹性容灾管理**：主要包括基于云原生的容灾备份、弹性扩容等功能。6G OSS 提供适用于网络云化、虚拟化、服务化演进需求的主动安全防护，提升网络故障管理响应能力，保障面对突发事件、灾难和攻击的弹性应对，确保网络和服务的可用性、可靠性以及连续性。

## 6.3 6G OSS 典型用例

### 案例一 6G 通感算性能联合优化

6G 网络是集通信、感知、计算为一体的信息系统，其业务承载在融合统一的 6G 硬件平台之上。因此，6G 网络管理系统需要对通信资源、计算资源和感知资源进行联合优化以满足用户 SLA 需求并实现网络运营效率最优。以智慧工厂中基于工业视觉的产品检测业务为例，6G 网络实时传输终端采集图像并通过内生算

力进行图像处理及检测。为了满足该业务中通信和计算融合的 SLA 需求，6G OSS 系统通过网络全自智运维管理功能调用意图管控功能分析业务传输计算模型和基站资源消耗趋势，并综合应用 6G OSS 内生的自动化、智能化和孪生化能力实现通信和计算资源最优编排调度方案并执行。

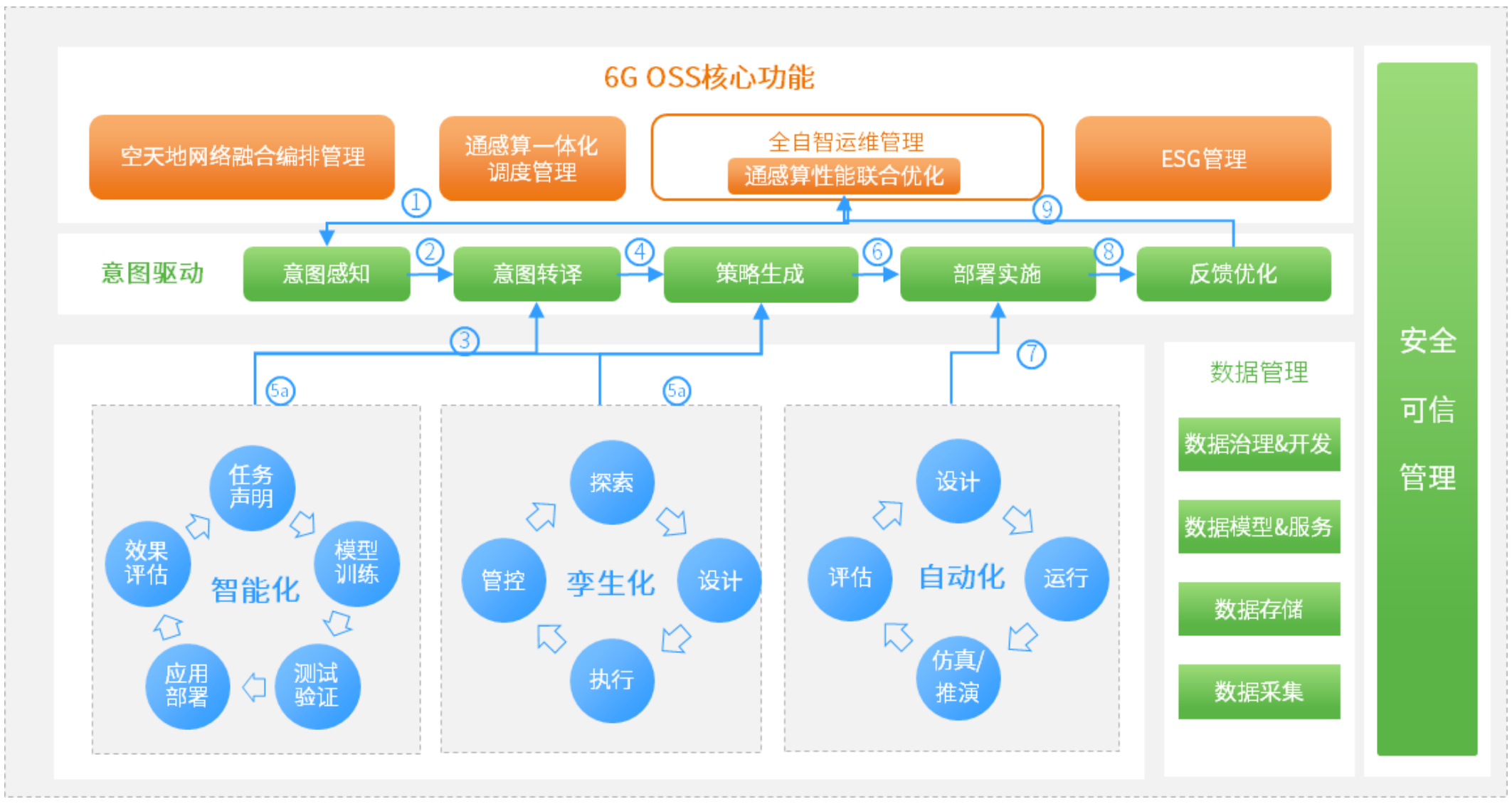

**图 6-6 6G 通感算性能联合优化流程**

如图 6-6 所示，在 6G OSS 系统中，基于工业视觉的产品检测业务所需的通算资源联合优化流程具体如下：

需求生成：根据基于工业视觉的产品检测业务，OSS 全自智运维管理功能中的通感算性能联合优化模块将生成通感算联合优化需求并下发至意图驱动功能；

意图感知：意图驱动功能中的意图感知模块通过调用智能化能力的自然语言处理、语义分析等模型分析输出具体的业务感知 SLA 需求并下发至意图转译模块；

意图转译：意图转译模块调用智能化能力的业务通感算性能映射模型和网络负荷预测模型，将业务 SLA 需求细化为业务传输计算模型和基站资源消耗趋势；

意图发送：意图转译模块将业务 SLA 需求细化为业务传输计算模型和基站资源消耗趋势发送到策略生成模块，以此作为策略学习的输入数据；

策略学习：策略生成模块调用孪生化能力构建产品检测业务相应网络区域的孪生场景，并通过智能化能力的网络资源配置强化学习算法在孪生场景中进行影响通感算性能的网络配置参数优化学习；

策略生成：策略生成模块根据基于孪生化和智能化的策略学习结果生成通感算网络参数调优策略并发送至部署实施模块；

部署实施：意图驱动功能应用自动化能力实现策略在现网的自动化执行，并将执行效果发送至反馈优化模块；

反馈优化：反馈优化模块监控策略部署实施的效果，并根据效果反馈对策略进行迭代优化；

策略上报：意图驱动功能将最终优化策略和优化效果反馈到全自智运维管理的通感算性能联合优化模块。

**案例二 6G 空天地网络业务开通**

为了支持无处不在的覆盖，满足用户高速移动的需求，空天地一体化网络有望成为未来 6G 网络的重要形态之一。通过包含地面基站、无人机、卫星等三维网络节点，6G 将为用户带来全球覆盖、万物互联的泛在通信服务。与传统陆地网络不同，这些异构通信节点构成了三维立体的网络空间，同时这些节点在业务支持能力、终端移动支持能力等方面存在较大差异，因此如何结合业务统筹考虑这些不同通信网络，最终实现优势互补满足业务需求将成为运营商考虑的重点，也是未来 6G OSS 的重要场景。可能的业务包括传统语音/数据业务、精准定位、图像识别、地质勘测、应急救灾、物流等。

在未来 6G 空天地一体网络中，用户业务发起后会有如传统地面蜂窝通信系统、高空通信系统、中低轨卫星、高轨卫星形成的立体网络承载，用户终端可同时使用多个星座资源并以加密隧道链路形式收发数据，这些星座的可提供速率差别极大，覆盖区域和服务时间窗口不同。可以根据具体的用户业务意图对其 SLA 进行分解，通过智能化能力实现对地面、天基网络的态势感知并形成包含实时及预测的资源视图、通过数字孪生化能力对不同资源承载时用户业务速率、时延等

关键性能指标进行仿真推演形成资源编排策略，最终借助自动化能力通过空天地融合编排管理实现端到端网络组网及业务开通及管理。

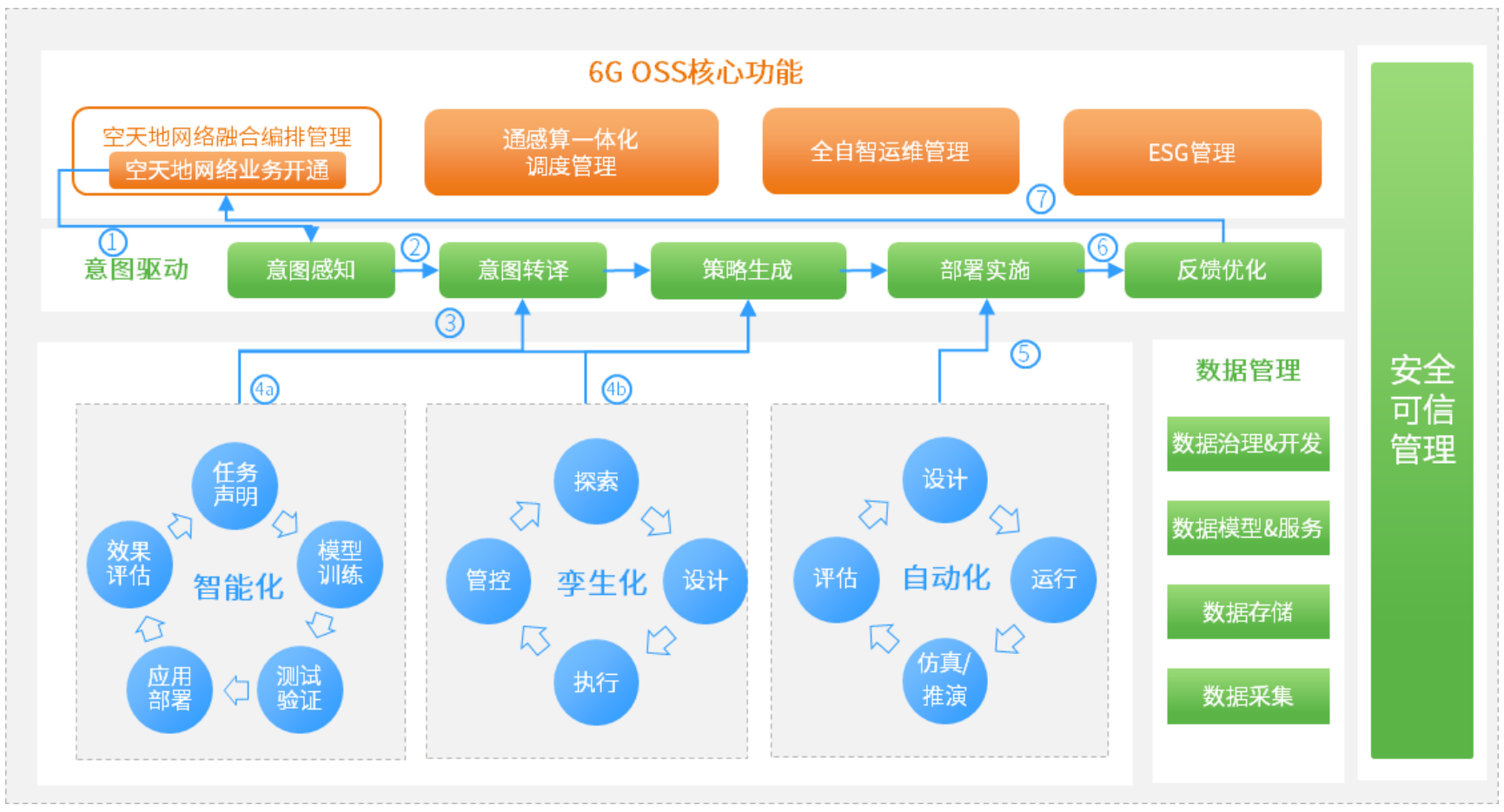

**图 6-7 6G 空天地网络业务开通流程**

如图 6-7 所示，在 6G OSS 系统中，6G 空天地网络业务开通流程具体如下：

需求生成：用户发起业务请求包含其 SLA 需求；

意图感知：意图驱动功能中的意图感知模块通过调用智能化能力的自然语言处理、语义分析等模型分析输出具体的业务感知 SLA 需求并下发至意图转译模块；

意图转译：意图转译模块调用智能化能力的空天地一体化态势感知能力对地面、高空、中低轨、高轨卫星资源进行分析预测，给出资源选择模型；

策略生成：策略生成模块调用孪生化能力及智能化能力给出最终用户需求的资源配置策略方案以及相关的 QoS 保障策略；

部署实施：意图驱动功能应用自动化能力实现策略在现网的自动化执行，并将执行效果发送至反馈优化模块；

反馈优化：反馈优化模块监控策略部署实施的效果，并根据效果反馈对策略进行迭代优化；

策略上报：意图驱动功能将最终资源编排策略及业务状态反馈到空天地网络融合编排管理模块及其他相关业务保障优化模块。

# 七. 6G OSS 的实现

在第四章和第五章中，我们重点分析了 6G OSS 系统的关键技术和功能架构，但是 6G OSS 系统的实现不是一蹴而就的，而是在现有网络的 OSS 系统基础上，逐步演进分阶段实现的。

## 7.1 5G OSS 现状

5G OSS 的主要功能可抽象为网络业务编排、网络资源管理、网络故障监控、网络性能与质量、网络体验分析、网络运维流程保障六个主要功能。

- **网络业务编排**：负责业务开通端到端高阶流程的设计与编排能力，统筹调度跨专业子流程，统一提供开通接口对接业务运营系统。实现流程的统一设计、运行及全程可视化管理功能。
- **网络资源管理**：负责运营商全专业网络资源数据管理、资源入网管理、资源调度管理、端到端网络资源拓扑视图等应用，提供各类资源服务[46]。
- **网络故障监控**：实现网络集中监控，提供网络监控开放能力，实现端到端业务感知监控以及以事件为中心的集中监控，并通过全自动服务化方式实现[47]。
- **网络性能与质量**：实现网络与业务质量的端到端分析及各类主题分析应用、网络性能集中监控分析，特别对如 VoNR 等跨域多专业业务实现质差识别、派单及闭环管理[48]。
- **网络体验分析**：实现对各类用户关键业务质量的体验定量分析，实现质差用户的识别、安抚及问题处理，并实现预防式运维[49,50]。
- **网络运维流程保障**：实现运维统一集中调度管理，提供自动派单等服务，涉及工单管理、代维管理、装维等应用，以及支撑上述应用的基础数据管理与统计分析功能。

与各类网络设备 OMC 相关的专业管理单元：实现专业内网络设备的自动操作维护和自动配置激活。

随着自智网络理念的不断深入，作为自智网络发展引擎的网络 OSS 系统也不断提升其自动化、智能化水平，随之而来目前主流运营商 5G OSS 体系中出现了一些通用化、平台化技术或系统，包括网络数据平台、AI 平台等，其主要功能如下：

- **网络数据平台**：网络数据平台是数字化转型及数据驱动的基础能力，负责网络管理领域各类数据的统一采集、存储。网络数据平台接入运营商全网全专业的网络与业务系统数据，在此基础之上对数据进行集中管理、集中存储与统一建模，同时面向运营商内部各类业务系统与应用提供数据同步、数据服务、租户入驻等多种形成的数据访问服务，支持批量数据和实时数据共享，可以满足运营商内不同系统的数据使用需求，为运营商网络全生命周期自动化网络提供数据基础保障。
- **网络 AI 平台**：网络 AI 能力的需求随着管理活动自动化程度的不断提升其作用越发明显，网络 AI 平台作为大规模智能服务的外挂式 AI 基础设施，向整个 OSS 系统提供体系化、工程化的 AI 技术能力。其可以单独或协同 NWDAF、SON 等功能实体，从各网元或网络系统统一收集数据信息，基于自智网络的各领域模型进行实时和离线推理，向各网络智能化应用注智赋能[51,52]。

## 7.2 6G OSS 的演进思路

6G OSS 系统是面向 6G 网络提供空天地、通感算一体化运维管理、赋能网络全域自动化、智能化、孪生化能力，并向社会提供 ESG 服务的内生安全和智慧的网络管理系统。如图 6-1 所示，6G OSS 系统的实现演进思路可分为智慧互联、协作赋能和融合一体三个阶段。

### 阶段一 分域互联型 6G OSS

在智慧互联阶段，6G OSS 系统将实现对 6G 空、天、地一体化新型网络的纳管，以及对通感算融合的新型业务的运营，同时它还将与现有如 4/5G 的现存网络之间连通。6G OSS 将采集解析不同制式网络的数据用以进行 6G 网络运维；通过系统间的互联交互实现多模网络数据的联合分析，进一步提升 OSS 的价值。同时，6G OSS 主要面向 6G 网络提供通感算资源的资源协同调度和管理，并实

现自动化、智能化、孪生化三种核心能力闭环与 6G OSS 系统各项功能之间的互联应用。

**阶段二 协作运维型 6G OSS**

在协作赋能阶段，6G OSS 系统将实现针对 6G 空天地网络和现有制式网络的不同网络子域之间的协作运维，并实现通感算业务之间的协作优化。6GOSS 的自动化、智能化和孪生化核心能力将与现有网络的运维系统及流程深度整合大幅提升存量网络的自智水平。

**阶段三 一体融合型 6G OSS**

在融合一体阶段，对于通感算资源深度融合的调度和管理、空天地一体网络业务选择与开通等新型业务，6G OSS 基于增强的意图驱动能力将实现 6G 网络和现有制式网络资源的综合调度管控，并实现基于 6G 空天地网络和现有制式网络融合架构的全域业务编排和网络运维，同时实现意图驱动的自动化、智能化、孪生化三种核心能力闭环的融合应用。

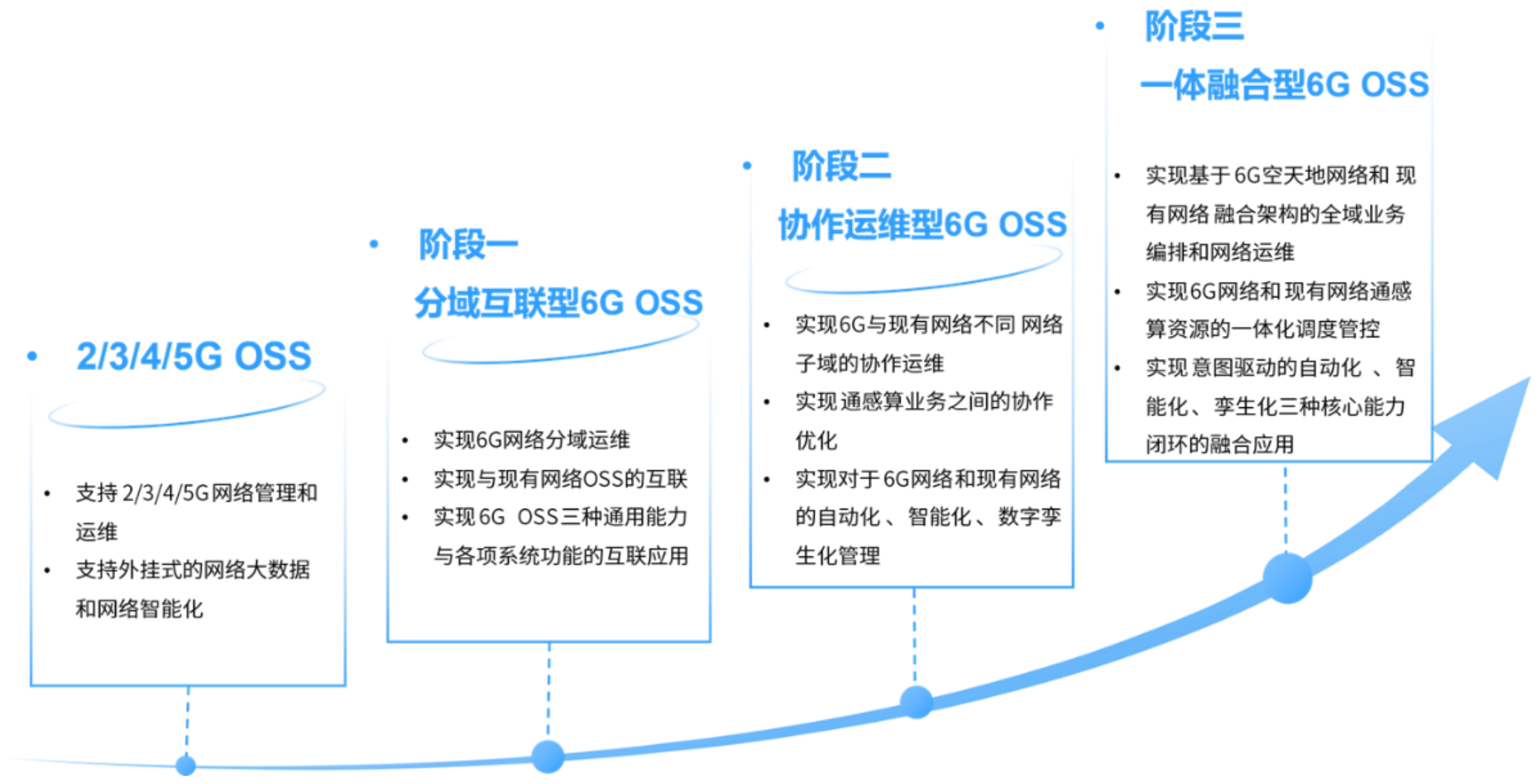

**图 7-1 6G OSS 系统的实现演进**

## 7.3 面向 6G OSS 的标准演进方向

伴随通信网络的代际演进，网络业务的复杂度不断提升，为提升实现网络运维及管理效率，以 OSS 系统为主要实现的网络管理架构在不同时期的标准化制定过程中也体现出不同演进重点。

### 3G：面向业务支持的网络管理架构

3G 系统的网络管理主要面向网络业务支持。国际电信联盟（ITU）于 20 世纪 80 年代后期引入的 TMN 框架。ITU-T M.3010 提出了 TMN 的逻辑分层架构（LLA）：网络元素层（NEL），网元管理层 EML，网络管理层 NML，业务管理层 SML，事务管理层 BML。ITU-T M.3400 规定了 TMN 五大管理功能域：故障管理、配置管理、计费管理、性能管理、安全管理( FCAPS )。TMN 应用领域非常广泛，涉及电信网及电信业务管理从业务预测到网络规划；从电信工程，系统安装到运行维护，网络组织；从业务控制和质量保证到电信企业的事物管理等。

90 年代中期 TMF 提出了面向电信行业的业务流程模型 TOM 模型（Telecom Operations Map），由于 TOM 模型缺少企业管理的内容，另外也缺少对互联网和电子商务催生的新业务支持，在 2001 年 TMF 提出了 eTOM 模型（enhanced TOM）来完善该模型。eTOM 中的过程管理域关注的焦点是在服务（Fulfillment），保障（Assurance）和 Billing & Revenue Management（计费），这三个组也被简称为“FAB”。eTOM 模型为整个行业提供了一个基础性的框架，成为事实上的行业标准和共同语言。

### 4G：面向虚拟化支持的网络管理架构

4G 时代的网络管理重点关注面向网络虚拟化的支持。为了加速部署新的网络服务，网络服务提供商和电信运营商积极拥抱网络功能虚拟化（NFV），从而可以逐步放弃笨重昂贵的专用网络设备。2012 年 10 月由 13 个运营商成立了欧洲通信标准协会 ETSI（European Telecommunications Standards Institute）的一个致力于推动“网络功能虚拟化的工作组（ETSI ISG NFV），通过负责开发制定电信网络的虚拟化架构，如 NFV MANO。

2014 年，ETSI 率先启动 MEC（Mobile Edge Computing，移动边缘计算）标准项目。这一项目组旨在移动网络边缘为应用开发商与内容提供商搭建一个云化计算与 IT 环境的服务平台，并通过该平台开放无线侧网络信息，实现高带宽、低时延业务支撑与本地管理。2016 年 ETSI 把 MEC 的接入方式，从蜂窝网络扩展到 WLAN 等其他接入方式，即把移动边缘计算的概念，扩展成为了新的 MEC（Multi-access Edge Computing，多接入边缘计算）。

**5G：面向智能化增强的网络管理架构**

随着 AI 技术的发展，5G 网络管理主要面向网络智能化增强。3GPP 在 5G 标准制定之初，就考虑将 AI 与大数据分析技术应用于 5G 网络。2017 初，R15 版本首次引入 NWDAF（Network Data Analytics Function，网络数据分析功能）网元，作为 5G 网络 AI+大数据引擎。2018 启动了“意图驱动的移动网络管理服务”并在 TR 28.812 中明确了意图驱动的网络管理服务的概念、自动化机制、应用场景以及描述意图的机制等。2020 年成立“自治网络分级（Autonomous Network Levels，ANL）”标准项目。此外，3GPP 还在 TS28.533 中定义了 MDAS（Management Data Analytics Service，管理数据分析服务）提供不同网络相关参数的数据分析，包括负载水平和/或资源利用率。

ETSI 于 2017 年成立业界首个网络智能化规范组——体验式网络智能行业规范小组（Experiential Networked Intelligence Industry Specification Group，ENI ISG），提出了利用 AI 和上下文感知策略来根据用户需求、环境状况和业务目标，通过自动化的服务提供、运营和保障等提升 5G 网络性能。同年 12 月，ISG ZSM（Zero Touch Network & Service Management）工作组成立，该工作组偏重无线和核心网，其标准化目标是端到端网络及服务进行自动化管理（如交付、部署、配置、维护和优化）。

2019 年，TM Forum 发起倡议并设立自智网络（Autonomous Networks)协作项目，旨在定义全自动化的零等待、零接触、零故障的电信网络，以支撑电信内部用户实现自配置、自修复、自优化、自演进的电信网络基础设施，探索并提供行业领先的端到端网络自动化方法论。

**5G Advanced：面向云原生服务化的网络管理架构**

随着云相关技术的不断发展和成熟，为了支持业务的快速上线、按需部署，5G 核心网经历了从 SDN/NFV，再到云原生的网络演进，面向 5G-Advance，网络服务化正由核心网向全网服务化演进。3GPP 已完成 5G 核心网服务化架构（SBA）以及基于服务化管理服务（MnS）的标准化研制，未来可以通过不断优化解耦，通过开放式服务化提升网络管理的效率。

随着各行业数字化转型的深入，面向数智驱动、泛在连接和虚实相生的未来网络新时代，2020 年 TM Forum 推出了 ODA（Open Digital Architecture）开放数字框架，通过提供标准化的云原生软件组件的方式助力运营商像搭积木一样搭建支持自动化运维的数字化 IT 系统，从而实现服务供应商及其供应商可以接受协同开发和跨组织的敏捷工作方式。

### 6G：面向通感算、空天地一体化的全新网络管理架构

基于以上网络管理相关架构的演进分析，面向未来 6G 通感算、空天地一体化的全新型网络架构，相应网络管理系统的设计也需要从过去打补丁式的单维能力增强或新增，向全栈 OSS 架构重构思考。

如图 7-2 所示，6G OSS 的网络管理核心功能需在原有传统网络业务支持的基础上需进一步新增面向空天地网络融合编排管理、通感算一体化调度管理、ESG 管理。通过数据管理以及意图管理等管理服务的引入，全面保障网络端到端服务质量和资源优化。在实现网络管理自动化、智能化之外，通过数字孪生技术进行网络仿真推演，为网络管理决策提供直观精确的孪生化的支撑。此外，6G OSS 还将从网络内生的角度提供安全可信以及 ESG 支持，全面保障 6G OSS 的可持续发展。

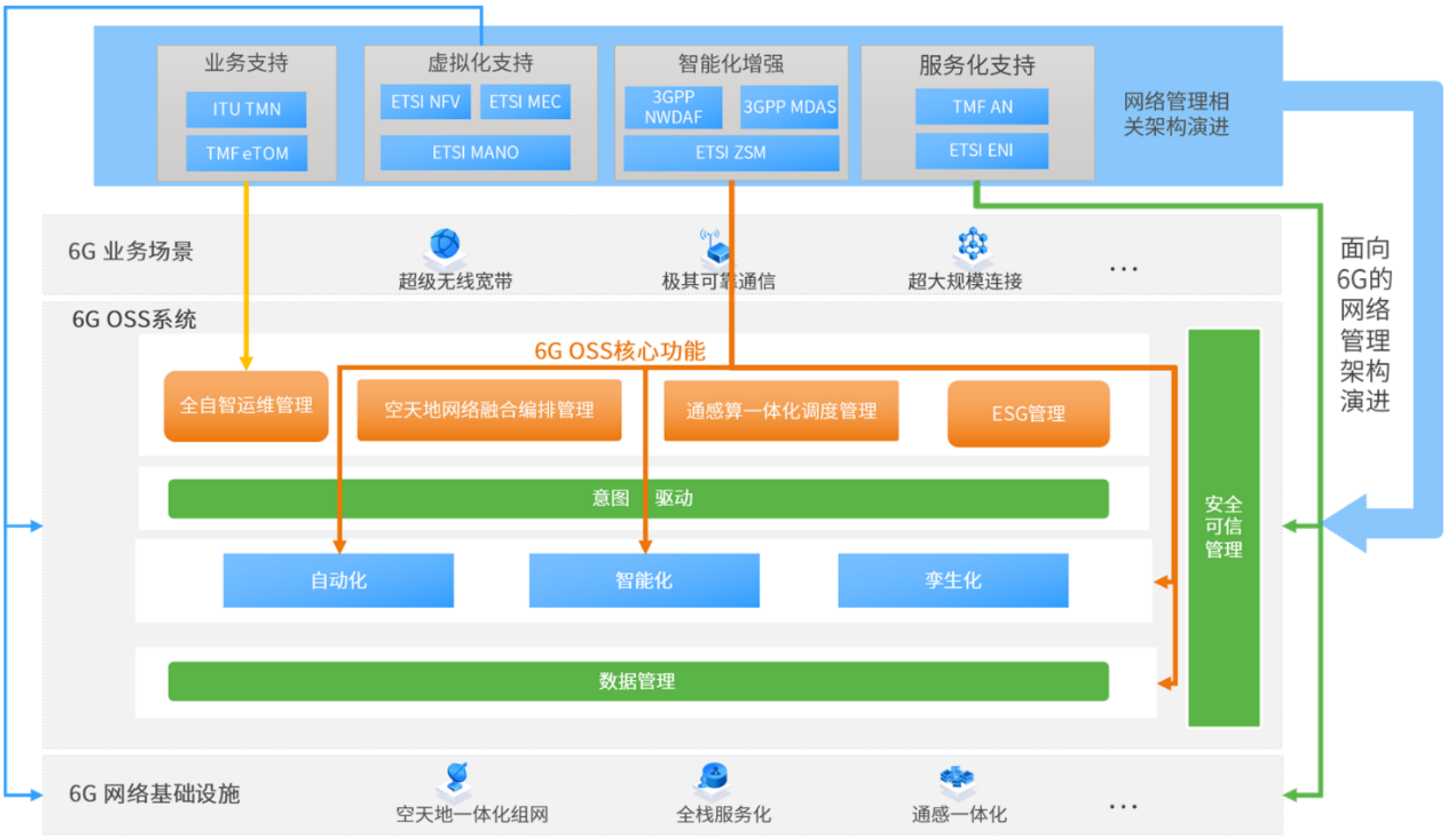

图 7-2 现有网络管理架构向 6G OSS 演进

# 八. 总结与展望

作为业界第一本前瞻性、系统性研究 6G OSS 的白皮书，本文从全球 6G 网络发展现状入手，全面分析了 6G 网络的典型业务场景、潜在关键技术和网络架构演进对于 OSS 的新需求；结合 6G 现状和 OSS 系统发展提出了 6G OSS 的总体愿景，研究了 6G OSS 系统的 12 项潜在关键技术，提出了由核心业务功能、三大管理功能和三大闭环核心能力组成的 6G OSS 系统功能架构，探讨向 6G OSS 系统演进的实现之路。

当前，业界对于 6G 系统架构和关键技术的研究尚处于预研阶段，对于 6G OSS 的研究刚刚起步，6G 基础设施建设与商用还是中远期目标。但是，从 6G OSS 的自动化、智能化、数字孪生化三大闭环核心能力看，业界已经具备了一定的研究基础和技术沉淀，并且这三项能力可以依托现有的 5G 网络智能化、算力网络、5G 专网等基础设施开展先行先试的技术研发和原型系统研制。同时，数字孪生等技术的率先发展也可以加速 6G 网络创新，降低行业研发成本和缩短研发周期。因此，面向 2030 年 6G 网络的商用目标，6G OSS 已具备了率先研发建设的基础条件和驱动力。

结合 6G OSS 的 5 大愿景，我们对于其发展路标做出如下展望：

在 2025 年， 完成面向分域互联型 6G OSS 的演进。此阶段 6G OSS 系统将支持面向 6G 空天地一体化架构的分域运维并与现有 4/5G 的现存网络之间实现连通，具备对于 6G 新无线技术的管理功能和基于算力内生网络的通算业务编排管理功能、并且初步具备数字孪生、内生 AI 和自动化三种能力的闭环及与 6G OSS 系统各项功能之间的互联应用；

在 2027 年， 完成面向协作运维型 6G OSS 的演进，实现“从网络智能化管理扩展到网络自动化、智能化、数字孪生化管理”和“构建安全可信的 6G OSS 体系”，并初步实现“从 5G 自智网络 L5 级向 6G OSS ready 演进”、“将环境、社会、治理（ESG）纳入 6G OSS 能力体系”的愿景目标。此阶段 6G OSS 系统将具备针对 6G 空天地网络和现有制式网络的不同网络子域之间的协作运维功能，并实

现通感算业务之间的协作优化功能，同时将实现 6G OSS 的数字孪生能力、内生 AI 能力和自动化能力与现存网络管理系统的协作应用。

在 2030 年，　完成面向一体融合型 6G OSS 的演进，全面实现 6G OSS“从网络单体/单域管理到空天地、通感算一体化管理演进”、“从 5G 自智网络 L5 级向 6G OSS ready 演进”等五大愿景目标。此阶段 6G OSS 系统将全面具备 12 项关键技术能力，实现 6G 网络和现有制式网络的资源一体化调度管控，6G 空天地网络和现有制式网络融合架构的全域业务编排和网络运维，同时实现基于增强型意图驱动的自动化、智能化、孪生化三种核心能力闭环的融合应用。

# 参考文献